\documentclass[12pt]{article}
\usepackage{graphicx}
\usepackage{amsmath}

\def\Re{{\cal R \mskip-4mu \lower.1ex \hbox{\it e}\,}}
\def\Im{{\cal I \mskip-5mu \lower.1ex \hbox{\it m}\,}}
\def\ie{{\it i.e.}}
\def\eg{{\it e.g.}}
\def\etc{{\it etc}}
\def\etal{{\it et al.}}

\def\sub#1{_{\lower.25ex\hbox{$\scriptstyle#1$}}}
\def\tev{\,{\rm TeV}}
\def\gev{\,{\rm GeV}}

\def\to{\rightarrow}

\def\be{\begin{equation}}
\def\ee{\end{equation}}
\def\bea{\begin{eqnarray}}
\def\eea{\end{eqnarray}}
\def\lum{\ifmmode {\cal L}\else ${\cal L}$\fi}
\def\inpb{\ifmmode {\rm pb}^{-1}\else ${\rm pb}^{-1}$\fi}
\def\infb{\ifmmode {\rm fb}^{-1}\else ${\rm fb}^{-1}$\fi}
\def\epem{\ifmmode e^+e^-\else $e^+e^-$\fi}
\def\ppb{\ifmmode \bar pp\else $\bar pp$\fi}
\def\mpl{\ifmmode \overline M_{Pl}\else $\overline M_{Pl}$\fi}
\def\cc{\ifmmode k/\overline M_{Pl}\else $k/\overline M_{Pl}$\fi}
\def\lpi{\ifmmode \Lambda_\pi\else $\Lambda_\pi$\fi}
\newskip\zatskip \zatskip=0pt plus0pt minus0pt
\def\matth{\mathsurround=0pt}
\def\lsim{\mathrel{\mathpalette\atversim<}}
\def\gsim{\mathrel{\mathpalette\atversim>}}
\def\atversim#1#2{\lower0.7ex\vbox{\baselineskip\zatskip\lineskip\zatskip
  \lineskiplimit 0pt\ialign{$\matth#1\hfil##\hfil$\crcr#2\crcr\sim\crcr}}}

\renewcommand{\thefootnote}{\fnsymbol{footnote}}

\begin{document} \begin{titlepage} 
\rightline{\vbox{\halign{&#\hfil\cr
&SLAC-PUB-10277\cr
&December 2003\cr}}}
\vspace{1in} 
\begin{center}

{\Large\bf  Higgsless Electroweak Symmetry Breaking in Warped Backgrounds:}
\vspace{12pt}
{\Large\bf Constraints and Signatures
\footnote{Work supported in part by the Department of
Energy, Contract DE-AC03-76SF00515}} 
\medskip
\end{center}

\centerline{H. Davoudiasl$^{1,a}$, J.L. Hewett$^{2,b}$,
B. Lillie$^{2, c}$, and T.G. Rizzo$^{2,d}$\footnote{e-mails:
$^a$hooman@ias.edu, $^b$hewett@slac.stanford.edu,
$^c$lillieb@slac.stanford.edu, and
$^d$rizzo@slac.stanford.edu}}
\vspace{8pt} \centerline{\it $^1$School of
Natural Sciences, Institute for Advanced Study,
Princeton, NJ 08540}
\vspace{8pt}
\centerline{\it $^2$Stanford Linear
Accelerator Center, Stanford, CA, 94309}

\vspace*{0.3cm}

\begin{abstract}

We examine the phenomenology of a 
warped 5-dimensional model based on SU(2)$_L \times$ SU(2)$_R \times$
U(1)$_{B-L}$ model which implements electroweak 
symmetry breaking through boundary conditions, without the presence
of a Higgs boson.  We use precision electroweak data to constrain 
the general parameter space of this model.  Our analysis includes 
independent $L$ and $R$ gauge couplings, radiatively 
induced UV boundary gauge kinetic terms, and all higher order corrections
from the curvature of the 5-d space.  We show that this setup can 
be brought into good agreement with the precision electroweak
data for typical values of the parameters.  However, we find that the 
entire range of model parameters leads to violation of perturbative 
unitarity in gauge boson scattering and hence this model is not a
reliable perturbative framework.  Assuming that unitarity 
can be restored in a modified version of this scenario, we consider the 
collider signatures.  It is found that new spin-1 states will be observed
at the LHC and measurement of their properties would identify
this model.  However, the spin-2 graviton Kaluza-Klein resonances, 
which are a hallmark of the Randall-Sundrum model, are too weakly 
coupled to be detected.

\end{abstract}

\renewcommand{\thefootnote}{\arabic{footnote}} \end{titlepage}

\section{Introduction}

After more than 30 years of experimental investigation, the
mechanism for electroweak Symmetry Breaking (EWSB) remains
unknown. The simplest picture of EWSB employs a scalar field, the
Higgs, whose vacuum expectation value provides masses for the
Standard Model (SM) $W^{\pm}$, $Z$ bosons, as well as for the 
fermions. Experiments have yet to find
this particle, even though generic expectations
place it within the reach of recent searches.  Direct searches
\cite{higgsmass} place the lower limit on the Higgs mass of
$m_h\gsim 114$ TeV, whereas a global fit to the precision
electroweak data set \cite{lepewwg} places the indirect upper bound of
$m_h<219$ GeV at 95\% CL.

On a more theoretical level, a weak scale Higgs scalar seems
unnatural, as its mass is typically expected to receive large
radiative corrections from UV physics.  Thus, a hierarchy problem
arises, as there seem to be much higher scales present in
Nature, such as the Planck scale of gravity, $\mpl \sim 10^{18}$ GeV.
This problem may be resolved by the addition of
new physics at the weak scale, such as
Higgs compositeness, strong dynamics (technicolor), or
supersymmetry. None of these proposals have been experimentally
verified, and they also suffer from various phenomenological
problems.

Over the past few years, the possibility of extra spatial 
dimensions has been exploited to
address the hierarchy conundrum. In particular, the warped
5-dimensional (5-d) Randall-Sundrum (RS) model \cite{RS}, which is based on
a truncated $AdS_5$ spacetime, offers a natural geometric setup for
explaining the size of the weak scale.  In this model, the weak scale
is generated exponentially from the curvature of the extra
dimensional space.  The AdS/CFT conjecture in
string theory \cite{Maldacena} suggests that the RS model is dual to a 4-d
strongly interacting field theory. The Higgs in the 5-d picture
is then identified with a dual 4-d composite scalar.  

It has been recently proposed \cite{CsakiI} that one
could use the boundary conditions of a 5-d flat space SU(2)$_L \times$
SU(2)$_R \times$ U(1)$_{B-L}$ theory to generate masses for $W^{\pm}$
and $Z$ bosons of the SM, in the absence of a Higgs scalar. 
This proposal predicted
unacceptably large deviations from precision EW data and seemed to be
excluded. However in Ref.\cite{CsakiII}, this Higgsless approach to EWSB 
was studied in the context of the RS geometry, and agreement with data
was much improved.  This can be understood from the fact that the
model contains a custodial SU(2) symmetry, as noted in 
Ref. \cite{raman} which is broken only by
terms of size of order the spatial variance 
of the bulk $W$ and $Z$ wavefunctions.
In the warped geometry, these wavefunctions are nearly flat over
most of the bulk, as opposed to the ${\cal O}$(1) spatial variance 
in the case of flat space.  

Using the AdS/CFT correspondence \cite{Maldacena}, one may
think of this proposal as a technicolor model without a Higgs
scalar.  This duality also addresses the improved agreement of the warped
model with data, since the global SU(2)$_L \times$ SU(2)$_R$ symmetry
in the bulk provides the equivalent of a 4-d custodial symmetry that
suppresses corrections to the EW observables.  Here, we note that even
though this construct is dual to some strong dynamics, the warped
5-d geometry could in principle provide a computationally controlled 
theory, with quantitative predictions.

In this paper, we study a 5-d Warped Higgsless Model (WHM), employing 
a set of parameters that is more general than those used in
the original model of Ref.\cite{CsakiII}.  In particular, we
allow for independent bulk gauge couplings to the $L$ and $R$ gauge
sectors, which is crucial in getting good agreement with the
precision EW data.  We also include the effects  
of UV boundary gauge kinetic terms, assuming that they are radiatively 
generated \cite{nomura}.  We
do not specify a mechanism for fermion mass generation, but adopt a
simple parametrization that could accommodate a large class of
possible scenarios.\footnote{
We note that a recent paper \cite{Csaki3} has proposed a mechanism
for generating the fermion masses geometrically by also employing boundary
conditions from the WHM configuration.}  In addition,
our analysis incorporates all higher order corrections from the
curvature of the 5-d space that were ignored in the initial
work\cite{CsakiII}.

We will demonstrate that with typical values for the model parameters, good
agreement with the precision EW data can be achieved.  
However, we have found that perturbative unitarity in $W_L^+W_L^-$ gauge boson
scattering is violated throughout the entire model parameter space. In
particular, for the region where the good agreement with precision
measurements is obtained, we find that unitarity
is violated at $\sqrt s\approx 2$ TeV, which is
below the mass of the new states studied by Csaki \etal\ \cite{CsakiI}.
We thus find that this model is not reliably predictive in its
present form.  However, assuming that
unitarity can be restored by an appropriate modification of this
scenario, \eg, with the inclusion of additional non-Higgs states,
we then consider the collider signatures which should be
present in any generic WHM.  In particular, we find that the gauge boson
Kaluza Klein (KK) excitations of the strong and electroweak sectors are
observable at the LHC.  However, it is unlikely that
the LHC experiments will be able to detect the spin-2 graviton KK
resonances which constitute the most distinct signature of the
conventional RS-based models.  

In the next section, we introduce our formalism and notation. We then
determine the couplings of the various KK towers to the SM fields in
Section 3.  Our predictions for the EW observables and the resulting
parameter space constraints are given in Section 4.
Unitarity is examined in Section 5 and the collider signatures of
the model are discussed in section 6.  Concluding remarks are
given in Section 7.

\section{Formalism and Notation}

In the analysis that follows, we will, for the most part, follow the
notation of Csaki \etal \cite{CsakiII} with some modifications 
that are necessary to make contact with our previous work \cite{dhr,dhrbt}.
For this reason we now review the RS metric in both notations.  In the 
original RS scheme (employed in our earlier work), the 5-d metric is 
given  by
\be
ds^2=g_{MN}dx^M dx^N = e^{-2\sigma}\eta_{\mu\nu}dx^\mu dx^\nu-dy^2\,,
\label{RSmetric}
\ee
with uppercase Roman indices extending over 5-dimensional 
space-time and Greek indices corresponding to 4-d.  
Here, $\sigma=k|y|=kr_c|\phi|$, with $r_c$ being the 
compactification radius, $k$ is the curvature scale associated with the 
5-d space, and $-\pi\leq\phi\leq\pi$ with $\phi$ parameterizing the 5th 
coordinate.  For numerical purposes we will take $kr_c=11.27$ throughout 
our analysis.  The geometrical setup contains 2 branes, one residing at 
$\phi=0$ (known as the Planck brane) and one at $\phi=\pi$ (the TeV 
brane), \ie, the branes are located at the boundaries of the 
5-dimensional Anti-de Sitter space.  We define the quantity 
$\Lambda_\pi\equiv \mpl e^{-\pi kr_c}$, which represents the scale of 
physical processes on the TeV brane.  In the scheme used 
in Ref. \cite{CsakiII}, this metric is rewritten as
\be 
ds^2 = \left( {R\over z} \right)^2 (\eta_{\mu\nu}dx^\mu dx^\nu - dz^2)\,,
\label{Csakimetric}
\ee
with $R\leq z\leq R'$.  Here, we see that the relationships $k=R^{-1}$, 
$R'=Re^{\pi kr_c}$, and $z=e^{ky}/k$ converts one form of the metric
to the other.  In this convention, the Planck (TeV) brane resides at 
$z=R\, (R')$.  It is important to note that the range $R\leq z\leq R'$ maps 
onto only {\it half} of the $-\pi\leq\phi\leq\pi$ interval.  When employing 
the Csaki \etal\ notation in what
follows, we will normalize our wavefunctions over twice the
$R\leq z\leq R'$  interval for consistency with our earlier work.

In the WHM, the gauge theory in the bulk is SU(3)$_C
\times $ SU(2)$_L\times$ SU(2)$_R\times$ U(1)$_{B-L}$ for which the
bulk action is given by
\be
S=\int d^4xdy\, \sqrt{-g}\sum_i {-1\over 4g^2_{5i}}F^i_{AB}F_i^{AB}\,,
\label{bulkaction}
\ee
where we have suppressed the group indices, $-g\equiv det(g_{MN})$,
the sum extends over the
four gauge groups, and $g_{5i}$ are the appropriate 5-d coupling 
constants.  Note that for generality, we allow for the possibility of 
$g_{5L}\neq g_{5R}$ in our analysis below.  The boundary conditions 
are chosen such that the gauge symmetry breaking chain SU(2)$_R\times$
U(1)$_{B-L}\to$ U(1)$_Y$ occurs at the Planck scale and subsequently
the gauge symmetry breaking 
SU(2)$_L\times$ U(1)$_Y\to$ U(1)$_{QED}$ takes place at the TeV scale.  
This hierarchical two-step breaking scheme is analogous to that of the usual 
breaking pattern of the conventional Left-Right Symmetric Model 
\cite{lrm}.  After the gauge symmetry is broken at the Planck scale,
a global SU(2)$_R \times$ SU(2)$_L$ symmetry remains in the
brane picture.  This global symmetry is broken on the TeV brane
to a diagonal group, SU(2)$_D$, which corresponds to the SU(2) custodial
symmetry present in the SM.  It is the presence of this custodial
symmetry which essentially preserves the tree-level value of unity 
for the $\rho$ parameter in this model.
Of the 7 generators present in the high-scale electroweak
sector, 3 are broken near \mpl, 3 are broken near the TeV scale, leaving
one generator for U(1)$_{QED}$ as in the SM. 
SU(3)$_C$, of course, remains unbroken and is simply
the 5-d analog of QCD.

In addition to the bulk action above, significant boundary (brane) 
terms can exist in this scenario \cite{nomura} which can be generated
via quantum contributions\footnote{There may be other brane terms
in the effective theory that can be important on both the IR and UV
branes.  In this treatment, we ignore such possible terms.}  \cite{georgi}. 
The only sizable effects arise at $y=0$, \ie,
on the Planck brane, due to the renormalization group evolution (RGE)
between the physical scales associated with the two branes, 
$\sim k$ and $\sim ke^{-\pi kr_c}$.  Since the gauge group below the
scale $k$ is simply SU(3)$_C\times$ SU(2)$_L\times$ U(1)$_Y$, only these
gauge fields will have brane localized kinetic terms, which
we may write as
\be
S_{brane}=\int d^4xdy\, \sqrt{-g}\, \delta(y)\left\{ -{1\over 4\tilde g^2_L}
F_L^{\mu\nu}F^L_{\mu\nu}-{1\over 4\tilde g^2_Y}F_Y^{\mu\nu}F^Y_{\mu\nu}
-{1\over 4\tilde g^2_s}F_C^{\mu\nu}F^C_{\mu\nu}\right\} \,,
\label{BLKTaction}
\ee
where
\be
{1\over \tilde g^2_i} = {\beta_i\over 8\pi^2}\ln({k\over ke^{-\pi kr_c}})
= {\beta_i\over 8\pi^2} \pi kr_c\,,
\label{tildeg}
\ee
for $i=L\,,Y\,,s$ and $\beta_i$ being the appropriate beta function.  If
only SM fields are present in the model, then $(\beta_L\,,\beta_Y\,,
\beta_s)=(-10/3\,,20/3\,,-7)$; we will assume these values in our numerical
analysis.  Note that due to the large logarithms, these coefficients
can be significant, of ${\cal O}(1)$ or larger, and will lead to
important effects as will be seen below.

In our earlier analysis \cite{dhrbt}, we introduced the notation
\be
{g^2_{5i}\over\tilde g^2_i}\, \equiv\, r_c c_i\, \equiv\, {2\delta_i\over k}\,,
\quad\quad\quad\quad\quad\quad (i=L\,,Y\,,s)
\label{dhrnote}
\ee
which is useful for quantifying the size of the brane kinetic terms.
Given the above relations for $1/\tilde g^2_i$, and the assumption that
only SM fields contribute to the beta functions, one can show that
$\delta_Y$ is not an
independent parameter, but is directly calculable.  (We will 
return to the case of 5-d QCD later.)  From Eq. (\ref{tildeg}) we have
\be
{1\over\tilde g^2_Y}=-2 {1\over\tilde g^2_L}\,,
\ee
where the factor of $-2$ arises from the ratio 
$\beta_Y/\beta_L$.  This leads to
\be
{g^2_{5Y}\over \tilde g^2_Y} = -2 {g^2_{5Y}\over\tilde g^2_L}\,.
\ee
Since SU(2)$_R\times$ U(1)$_{B-L}\to$ U(1)$_Y$, we have the relations
\be
{1\over g^2_{5Y}} = {1\over g^2_{5R}} + {1\over g^2_{5B}}\,,
\ee
which we can write as
\be
{g^2_{5L}\over g^2_{5Y}}={1\over\kappa^2}+{1\over\lambda^2}\,,
\ee
by introducing the notation
\be
\kappa\equiv {g_{5R}\over g_{5L}}\,,
\quad\quad\quad\quad \lambda\equiv {g_{5B}\over g_{5L}}\,.
\ee
Solving the above for $g^2_{5Y}$, we obtain
\be
{g^2_{5Y}\over \tilde g^2_Y} = -2 {\lambda^2\kappa^2\over
\lambda^2+\kappa^2}{g^2_{5L}\over\tilde g^2_L}\,,
\ee
which yields
\be
\delta_Y = -2 {\lambda^2\kappa^2\over\lambda^2+\kappa^2}
\delta_L\,.
\ee
As we will see below, the value of $\lambda$ will be determined
by the $M_{W,Z}$ mass relationship while $\kappa$ will remain
a free parameter confined to a constrained region.

The following set of
boundary conditions generate the symmetry breaking pattern
discussed above (note that we suppress the Minkowski indices):

\noindent On the TeV brane at $z=R'\,\,\, (y=\pi r_c)$ one has
\bea
\partial_z(g_{5R}A^a_L+g_{5L}A^a_R) & = & 0\,;\quad\quad\quad
\partial_z A^a_C=0\,; \nonumber\\
g_{5L}A^a_L-g_{5R}A^a_R & = & 0\,;\quad\quad\quad
\partial_z B=0\,;\\
g_{5L}A_5^{La}+g_{5R}A_5^{Ra} & = & 0\,;\quad\quad\quad
B_5=0\,;\nonumber\\
\partial_z (g_{5R}A_5^{La}-g_{5L}A_5^{Ra}) & = & 0\,,\nonumber
\label{bctev}
\eea
with $A^a_{L(R)}$ or $A_C^a$ being one of the SU(2)$_{L(R)}$ or
SU(3)$_C$ fields with gauge index $a$, and $B$ being the
corresponding U(1)$_{B-L}$ field.

\noindent On the Planck brane at $z=R\,\,\, (y=0)$ the boundary conditions are
\bea
\partial_z A_L^a & = & -\delta_Lx_n^2k\epsilon^2A_L^a\,;
\quad\quad\quad \partial_z A_C^a=-\delta_sx_n^2k\epsilon^2A_C^a\,;
\nonumber\\
A_R^{1,2}& = & 0\,;\quad\quad\quad g_{5B}B-g_{5R}A_R^3=0\,;\\
\partial_z [ g_{5B}A_R^3 & + & g_{5R}B]=-\delta_Yx_n^2k\epsilon^2
[g_{5B}A_R^3+g_{5R}B]\,;\nonumber\\
A_5^{La} & = & 0\,;\quad\quad A_5^{Ra}=0\,;\quad\quad B_5=0\,,\nonumber
\label{bcplanck}
\eea
where $\epsilon=e^{-\pi kr_c}$, and $m_n=x_nk\epsilon$ is the
mass of the n$^{th}$ gauge KK state.  For the remainder of this
paper, we will work in the unitary gauge, where the fifth
components of the gauge fields are zero.

Recalling the breaking
pattern for SU(2)$_R\times$ U(1)$_{B-L}\to$ U(1)$_Y$, we 
introduce the fields
\bea
Y & = & {g_{5R}B+g_{5B}A_R^3\over\sqrt{g^2_{5R}+g^2_{5B}}}\,,
\nonumber\\
\zeta & = & {g_{5B}B-g_{5R}A_R^3\over\sqrt{g^2_{5R}+g^2_{5B}}}\,,
\eea
and identify $Y$ with the usual hypercharge field.  In that case,
the  boundary condition on the third line in Eq. (15) can be
written more simply as
\be
\partial_z Y = -\delta_Yx_n^2k\epsilon^2Y\,.
\ee

The KK decomposition we use is essentially that of Csaki \etal, but
allowing for $g_{5L}\neq g_{5R}$ and is expanded to include the 
SU(3)$_C$ group:
\bea
B(x,z) & = & \alpha_B\gamma(x)+\sum \chi^B_k(z)Z^{(k)}(x)\,,\nonumber\\
A_L^3(x,z) & = & \alpha_L\gamma(x)+\sum \chi^{L^3}_k(z)Z^{(k)}(x)
\,,\nonumber\\
A_R^3(x,z) & = & \alpha_R\gamma(x)+\sum \chi^{R^3}_k(z)Z^{(k)}(x)
\,,\nonumber\\
A_L^\pm(x,z) & = & \sum \chi^{L^\pm}_k(z)W^{(k)\pm}(x)\,,\\
A_R^\pm(x,z) & = & \sum \chi^{R^\pm}_k(z)W^{(k)\pm}(x)\,,\nonumber\\
A_C(x,z) & = & \alpha_g g(x)+\sum \chi^g_k(z)g^{(k)}(x)\,,\nonumber
\eea
where we have again suppressed the Lorentz indices and the sum extends over
the KK tower states, $k=1...\infty$.  Here, $\gamma(g)$ is the massless
photon(gluon) field and $\alpha_{B,L,R,g}$ are numerical
constants which are determined from the boundary conditions.  Note that
since the photon and gluon zero-mode states are massless, their 
wavefunctions are $z$-independent, \ie, they are `flat' in $z$.  The
wavefunctions $\chi^A_k(z)$ take the form
\be
\chi^A_k(z)=z(a_A^kJ_1(m_kz)+b_A^kY_1(m_kz))\,,
\ee
with $J_1\,,Y_1$ being first-order Bessel functions with the
coefficients $a_A^k\,,
b_A^k$ and the KK masses $m_k$ to be determined by the boundary
conditions as we now discuss.

Let us first consider the case of the charged gauge boson sector.
We first introduce the notation,
\bea
R_i & \equiv & Y_i(x_n^W\epsilon)/J_i(x_n^W\epsilon)\,,\nonumber\\
\tilde R_i & \equiv & Y_i(x_n^W)/J_i(x_n^W)\,,
\eea
where $x_n^Wk\epsilon$ are the masses of the $W^\pm$ KK tower
states.  Expressions for the coefficients $b^\pm_{L,R}\,,
a^\pm_{R}$ in terms of $a^\pm_L$ can easily be obtained via
the boundary conditions; we find (dropping the KK index for
convenience)
\bea
b^\pm_L & = & -{a_L^\pm\over R_0}X_L\,,\nonumber\\
b_R^\pm & = & -{a_R^\pm\over R_1}\,,\\
a_R^\pm & = & -\kappa {(1-X_L\tilde R_0/R_0)\over (1-\tilde R_0/R_1)}
a_L^\pm\,,\nonumber
\eea
where $a_L^\pm$ will be determined by the wavefunction normalization and
\be
X_L\equiv {1+\delta_Lx_n^W\epsilon R_W\over 1+\delta_L
x_n^W\epsilon R_1R_W/R_0}\,,
\ee
with $R_W\equiv J_1(x_n^W\epsilon)/J_0(x_n^W\epsilon)$.
The masses of the KK states can then be determined and 
are explicitly given by the root equation
\be
(R_1-\tilde R_0)(R_0-X_L\tilde R_1)+\kappa^2(R_1-\tilde R_1)
(R_0-X_L\tilde R_0) = 0\,.
\ee
Note that for $g_{5L}=g_{5R}$, \ie, $\kappa=1$, and in the absence
of boundary terms ($\delta_L=0$, $X_L=1$), this expression reduces 
to that obtained by Csaki \etal\ \cite{CsakiII}.  
We will return to a study of the roots
and corresponding gauge KK masses in the next section.

We now turn to the neutral electroweak sector and first consider the
massive tower states.  The boundary conditions yield (where $R_i\,,
\tilde R_i$ are defined as above with $W\to Z$)
\bea
b_L & = & -{a_LX_L\over R_0}\,,\nonumber\\
b_R & = & -a_L{(1-X_L\tilde R_1/R_0)+\kappa^2(1-X_L\tilde R_0/R_0)
\over \kappa(\tilde R_0-\tilde R_1)}\,,\nonumber\\
a_R & = & -\kappa a_L(1-X_L\tilde R_0/R_0)-b_R\tilde R_0\,,\\
b_B & = & -{a_B\over\tilde R_0}\,,\nonumber\\
a_B & = & -{\lambda a_L(X_Ya_R/a_L+R_0b_R/a_L)\over
\kappa(X_Y - R_0/\tilde R_0)}\,,\nonumber
\eea
where now $a_L$ is determined via normalization and we have 
defined
\be
X_{L,Y}\equiv {1+\delta_{L,Y}x_n^Z\epsilon R_Z\over
1+\delta_{L,Y}x_n^Z\epsilon R_ZR_1/R_0}\,,
\ee
with $R_Z\equiv J_1(x_n^Z\epsilon)/J_0(x_n^Z\epsilon)$.
The root equation for the neutral KK tower masses is then,
\bea
& & -\lambda^2 (\tilde R_0-R_1) 
\left\{  \kappa^2X_Y(\tilde R_0-
\tilde R_1)(R_0-X_L-\tilde R_0)+(R_0-X_L\tilde R_1)(R_0-X_Y
\tilde R_0)\right.\nonumber\\
& + & \left. \kappa^2(R_0-X_L\tilde R_0)(R_0-X_Y\tilde R_0)
\right\}\nonumber\\
& + & \kappa^2(R_0-X_Y\tilde R_0)\left\{ \kappa^2(\tilde R_0 -
\tilde R_1)(R_0-X_L\tilde R_0)+(R_1-\tilde R_0)(R_0-X_L
\tilde R_1)\right.\nonumber\\
& + & \left. \kappa^2(R_0-X_L\tilde R_0)(R_1-\tilde R_0)\right\} = 0\,.
\eea
Note that unlike the case where brane terms are neglected, this
equation does not factorize into a pair of KK towers associated
with the $\gamma$ and $Z$.  In fact, as we will see below, the
$\gamma$ and $Z$ tower states are highly mixed and do not 
simply appear to be more massive copies of the SM photon and $Z$
boson.

Turning to the case of the zero-mode photon, the fact that its
wavefunction is constant in $z$ trivializes all but two of the
boundary conditions from which the coefficients in Eq. (18)
may be obtained:
\be
\alpha_R=\alpha_L/\kappa\,,\quad\quad\quad\quad \alpha_B=\alpha_L/
\lambda\,,
\ee
with $\alpha_L$ to be determined via normalization of the massless
photon field.

For the remaining case of SU(3)$_C$, we see that $\alpha_g$ is 
determined via the normalization of the massless gluon field 
and the two boundary conditions lead to the single relation
\be
b_s=-{a_sX_s\over R_0}\,,
\ee
where $X_s$ is defined from Eq. (22) with 
$x_n^{W}\to x_n^g$, $\delta_L\to\delta_s$. 
The mass spectrum of the gluon excitations are then
given by the simple relation
\be
R_0-X_s\tilde R_0=0\,.
\ee
As before, $a_s$ will be determined via the normalization
conditions in the next section.

\section{Determination of the KK Mass Spectrum and Couplings}

\begin{figure}[htbp]
\centerline{
\includegraphics[width=9cm,angle=90]{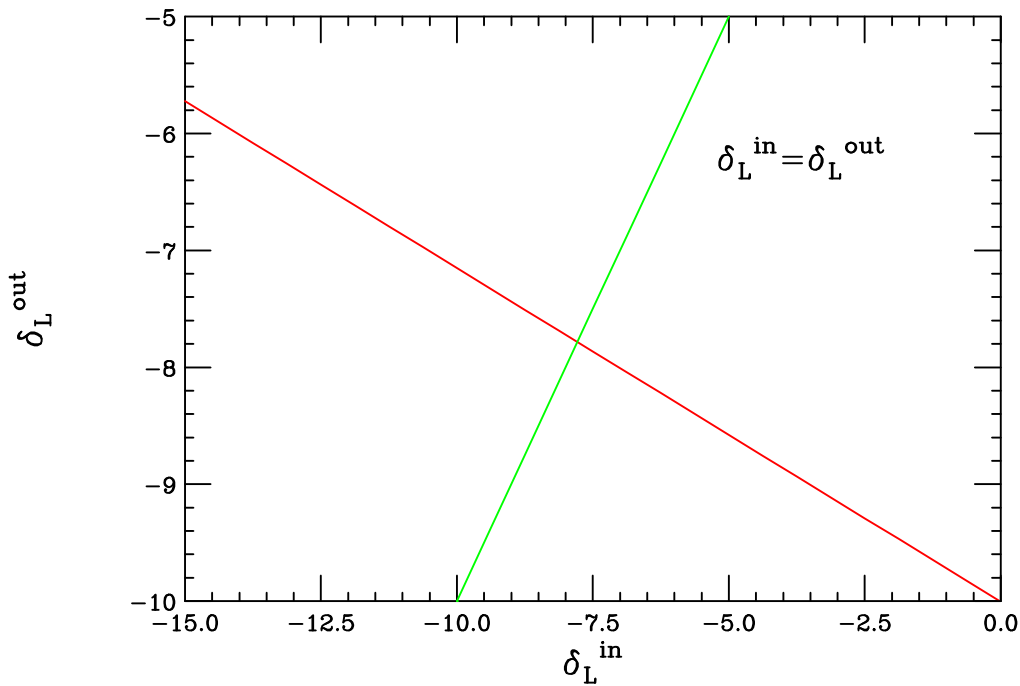}}
\vspace*{0.1cm}
\caption{The value of $\delta_L
(=\delta_L^{out})$ calculated via the procedure described in the text
as a function of the input value (red curve).  The curve corresponding
to $\delta_L^{in}=\delta_L^{out}$ is also shown (green curve); the solution
lies at the intersection of the two curves.  Here, $\kappa=3$ is
assumed.}
\label{fig1}
\end{figure}

In this section, we solve the various root equations to determine
the mass spectrum and couplings of the KK sector.
A priori, it would seem that the parameters $\kappa\,,\lambda$,
and $\delta_{L}$ are completely arbitrary, but as we will see, 
some of them are determined by data.  We first
consider the $W^\pm$ KK tower.  In this case, the root
equation depends on $\kappa$ and $\delta_L$, and although $1/\tilde
g^2_L$ is known, the ratio $g^2_{5L}/\tilde g^2_L$, which gives
$\delta_L$, is not.  However, $\delta_L$ is not arbitrary and can
be determined from the measured value of the Fermi constant
in a self-consistent manner as follows.  Our approach
is:  ($i$) we choose a value of $\kappa$ and an input value of $\delta_L
(=\delta_L^{in})$ and then calculate the roots $x_n^W$ using
Eq. (22).  Since $m_{W_1}=M_W=x_1^Wk\epsilon$ is identified with the 
physical $W^\pm$ state observed in experiment and is thus known, 
this fixes $k\epsilon$ so that the masses of all the KK excitations $m_{W_n}$
can be determined.  ($ii$)  Now that the values of the $x_n^W$ are 
known, the coefficients $a_R^\pm\,, b^\pm_{L,R}$ of
the wavefunctions are also calculable; 
this allows us to determine the couplings of the KK states
to the SM fermions.  These are of the form
\be
g^2_{W_n}=N_n{g_{5L}^2\over 2\pi r_c}\,,
\ee
where the coefficients $N_n$ are computed below.  ($iii$) We observe that
the above equation can be rewritten to give $g^2_{5L}$ provided
that $g^2_{W_1}$, which corresponds to the usual $W$ boson coupling,
is known.  We find
\be
g^2_{5L}=2\pi r_c g^2_{W_1}/N_1\,,
\ee
so that
\be
\delta_L^{out} = \pi kr_c{g^2_{W_1}\over N_1\tilde g^2_L}\,.
\ee
($iv$) Next we must examine whether $\delta_L^{out}=\delta_L^{in}$
as a test of consistency.
We recall from $\mu$-decay that at tree-level,
\be
{8G_F\over\sqrt 2}=\sum_{n=1}^\infty {g^2_{W_n}\over m^2_{W_n}}
={g^2_{W_1}\over M_W^2}\sum_{n=1}^\infty
{N_n\over N_1(x^W_n/x^W_1)^2}\,,
\ee
with all the quantities in this equation 
being known except $g^2_{W_1}$.  Hence
solving for $g^2_{W_1}$ and inserting this result into the previous
equation, we obtain a calculable expression for $\delta_L^{out}$,
\be
\delta_L^{out}={\pi kr_c\over N_1\tilde g^2_L}{8G_FM_W^2\over
\sqrt 2}\left[ \sum_{n=1}^\infty {N_n\over N_1(x_n/x_1)^2}\right]^{-1}\,.
\ee
If $\delta_L^{out}\neq\delta_L^{in}$ for a fixed value of $\kappa$,
we perform another search until convergence is obtained.  When we
have found a consistent solution (\ie, $\delta_L^{in}=\delta_L^{out}$),
$g^2_{W_n}$ and $g^2_{5L}$ are determined absolutely as a function 
of $\kappa$.  We would expect $\delta_L$ to only weakly depend on 
$\kappa$ since this dependence vanishes at
zeroth-order in $1/\pi kr_c\simeq 1/35$.  Here, we emphasize
the importance of the $\mu$-decay constraint in obtaining this
result, as  $G_F$ determines the absolute strength
of the coupling $g_{W_1}$, thus providing a reference to which
all others can be scaled.    Fig. \ref{fig1} displays
the value of $\delta_L^{out}$ versus $\delta_L^{in}$ for $\kappa=3$,
for demonstration, and we see that a unique solution is obtained only when
$\delta_L\simeq-7.8$; we find similar results for other values
of $\kappa$ near unity.

The parameter $\kappa$ is bounded from 
below as can be seen from Eqs. (10)
and (11) of Ref. \cite{nomura} which apply at lowest order in
$1/\pi kr_c$,
\be
c_w^{-2}={M_Z^2\over M_W^2} = {\kappa^2\lambda^2(1+D_L)
+(\kappa^2+\lambda^2)(1+D_Y)\over (\kappa^2+\lambda^2)(1+D_Y)}\,,
\ee
where $D_{L,Y}=\delta_{L,Y}/\pi kr_c$, and $c_w=\cos\theta_w$ where
$\theta_w$ is the weak mixing angle.    Solving for $\lambda^2$,
using $D_Y=-2D_L\kappa^2\lambda^2/(\kappa^2+\lambda^2)$, and
demanding that $\lambda^2>0$, we obtain the bound
\be
\kappa^2 > {c_w^{-2}-1\over 1+D_L(2c_w^{-2}-1)} \,.
\ee
With $\delta_L\simeq -7.8$ and $c_w^2\simeq 0.78$, this
implies the constraint
\be
\kappa\gsim 0.66\,.
\ee  
Although there will be corrections to this
result from terms of order $1/\pi kr_c$, we expect these to be no
more than a few percent.  To be concrete, we will assume that
$\kappa\geq 0.75$ in our analysis.

It is also possible to obtain an approximate upper bound on
$\kappa$ based on perturbativity arguments, as the typical
4-d couplings $g^2_{4R}\equiv g^2_{5R}/2\pi r_c$ cannot become
too strong.  A short analysis leads to the constraint that
$\kappa\lsim 4$.  To be specific,
we will thus limit ourselves to the range $0.75\leq \kappa\leq 4$
in our study.  This agrees with our expectations that 
on general grounds, the values of
$g_{5R}$ and $g_{5L}$ should not be too different, implying that
$\kappa\sim 1$.

Next, in order to define the KK couplings to 
SM fields, we need to discuss the localization of the SM fermions.
(Note that we define the strength of the `weak coupling' via the 
interaction of the SM $W^\pm$ boson and fermions.)  In the original
analysis of the WHM \cite{CsakiI,nomura}, the SM fermions were all localized
on the Planck brane; for further model-building purposes this need 
not be so \cite{Csaki3}.  However, it is well-known that if the
fermions are localized close to the Planck brane their gauge
couplings can be well approximated by the purely Planck brane values
\cite{dhr3,pomgher}.  We have checked that the gauge field wavefunctions
are reasonably flat for fermions localized with $\nu\lsim -0.6$,
where the quantity $\nu$ is as defined in Ref. \cite{dhr3}.
For simplicity, we thus make this assumption below.
Under this assumption, the covariant derivative acting on these
fields is given by 
\be
D^\mu=\partial^\mu+ig_{5L}T_LA_L^\mu+ig_{5R}T_RA_R^\mu+
ig_{5B}{B-L\over 2}B^\mu+ig_{5s}T_sA_C^\mu\,.
\ee

In this case, following Csaki \etal\ \cite{CsakiII} and 
Nomura \cite{nomura}, the couplings of 
the $W_n$ KK states to the SM fermions are given by
\be
g^2_{W_n}=\Omega^2{g^2_{5L}|\chi_n^{L^\pm}(R)|^2\over
N_{W_n}}\,,
\ee
with
\be
N_{W_n}=\int_R^{R'}dz\,{R\over z}\left\{ |\chi_n^{L^\pm}(z)|^2
[2+c_Lr_c\delta(z-R)]+2|\chi_n^{R^\pm}(z)|^2\right\} \,,
\ee
where the relative factors of 2 arise from the interval extension
discussed in the previous section.  The coefficient $\Omega$
is determined numerically via the self-consistency procedure described 
above, which demands that for $n=1$ (\ie. the SM $W$ boson),
we recover the usual SM coupling $g^2_{W_1}=g^2_{SM}$.  Thus,
the $W$ boson coupling automatically retains its known value by construction
when we identify $W_1^\pm\equiv W^\pm_{SM}$ as the experimentally
observed state (and correspondingly $m_{W_1}=M_W$ through the use
of $G_F$).  Using $M_W$ and $g^2_{SM}$ from experiment, we thus
can determine the masses and couplings of all the higher KK
modes.  Here, we assume the LEPEWWG \cite{lepewwg} central 
values of the SM gauge boson masses, $M_W=80.426\gev$ and
$M_Z=91.1875\gev$ in our analysis.

\begin{figure}[htbp]
\centerline{
\includegraphics[width=9cm,angle=90]{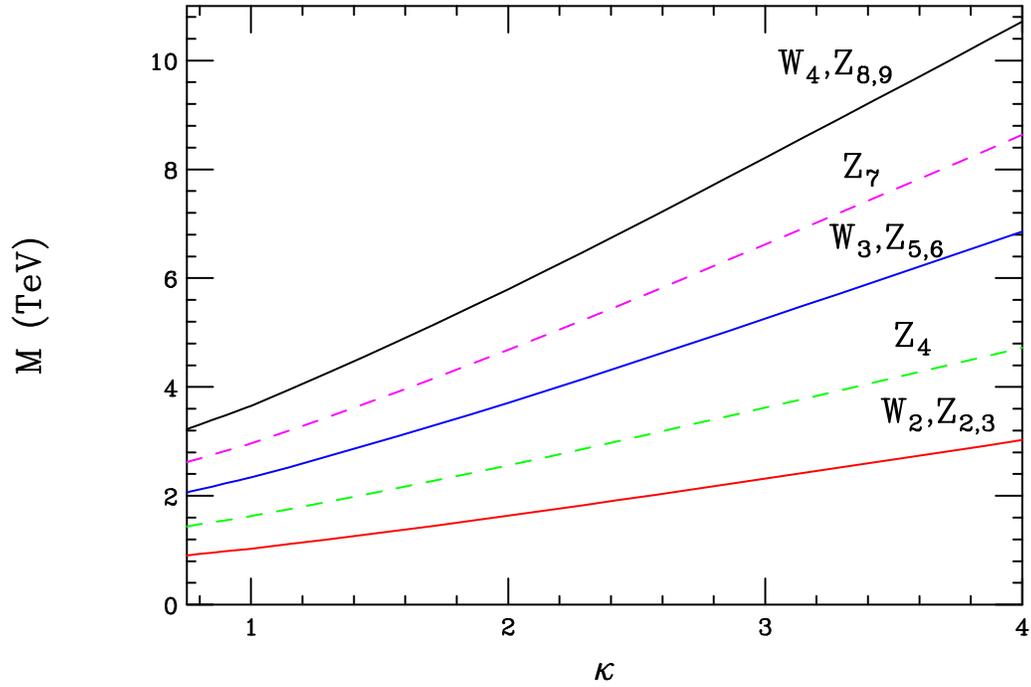}}
\vspace*{0.1cm}
\caption{Masses of the electroweak gauge KK excitations as a function
of $\kappa$.  The solid curves correspond to the $W^\pm$ states, while
the $Z$ KK excitations correspond to both the solid and dashed curves
as labeled.
In the latter case, the solid curves correspond to the almost doubly
degenerate states.}
\label{figmasses}
\end{figure}

\setlength{\unitlength}{1.5cm}
\begin{figure}[tbh]
\begin{centering}
\begin{picture}(10,3)
\thicklines
\put(0.5,.52){\line(1,0){3}}
\put(4.5,.5){\line(1,0){3}}
\put(4.5,.54){\line(1,0){3}}
\put(4.5,1){\line(1,0){3}}
\put(0.5,1.52){\line(1,0){3}}
\put(4.5,1.5){\line(1,0){3}}
\put(4.5,1.54){\line(1,0){3}}
\put(4.5,2){\line(1,0){3}}
\put(0.5,2.52){\line(1,0){3}}
\put(4.5,2.5){\line(1,0){3}}
\put(4.5,2.54){\line(1,0){3}}
\put(2,0){W}
\put(6,0){Z}
\end{picture}
\caption{Schematic comparison of the $W^\pm$ and $Z$ KK mass spectra showing
that the $W^\pm$ KK states have masses almost identical to those of the
degenerate pair of $Z$ KK excitations.}
\label{massspec}
\end{centering}
\end{figure}
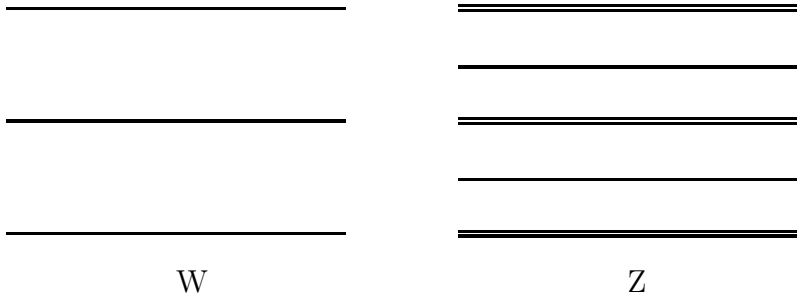

Figure \ref{figmasses} displays the masses of the first few $W$ KK
excitations as a function of $\kappa$.  We see that the masses
grow reasonably rapidly as $\kappa$ increases and can be quite
heavy.  For example, for $\kappa=3$, the first $W^\pm$ excitation
above the SM-like $W$ boson, $W_2^\pm$, has a mass of $\simeq
2.32$ TeV.  The masses of the higher KK states are approximately
given by the root relation $x_n^W=x_2^W+(n-2)\pi$.
The coupling strength of the gauge KK excitations are small
relative to those for the $W^\pm$ and
decrease rapidly as the KK mode number increases.  {\it E.g.},
the first $W^\pm$ KK excitation has a fermionic coupling of
only $g^2_{W_2}\simeq 0.0431g^2_{SM}$.  As we will see below,
this will have important implications in the consideration of
unitarity violation in $W_LW_L$ scattering.
 
We now turn to the neutral KK states and first discuss their
mass spectrum.  We will refer to these states as $Z_n$, but they
are KK excitations of both the $\gamma$ and $Z$ and are mixtures
thereof.  In our analysis, we will force the $W$ and 
$Z$ bosons to have the correct masses, \ie, those given by
experiment, and will also make use of the on-shell definition
of the weak mixing angle.
\be
\cos^2\theta_w^{os}\equiv {M_W^2\over M_Z^2}\,.
\ee
Thus, determining the roots $x_1^W(\kappa)$ from the analysis
discussed above, yields
\be
x_1^Z(\kappa)={M_Z\over M_W}x_1^W(\kappa)\,,
\ee
with the ratio $M_Z/M_W$ taken as exactly known.  Note that we identify
the lightest massive neutral KK state with the $Z$ boson
observed at LEP/SLC.   In order to solve the $Z_n$ eigenvalue
equation (26), we input our chosen value of $\kappa$ and our determined
value of $\delta_L$ from which we can obtain $\delta_Y$;  
$\lambda$ remains an independent variable, but we pick its value
in order to obtain the correct root $x_1^Z$ above.  Once this is
accomplished, all of the electroweak parameters in the model
(except $\kappa$) are completely determined, in particular, the
$Z_n$ KK tower masses and the wavefunction coefficients $a_{R,B}$
and $b_{L,R,B}$ of Eq. (19).

The masses of the $Z_n$ KK tower states have an unusual behavior;
there is a repeating pattern of a pair of almost degenerate states, 
followed by a single state, \eg, the
states $Z_2$ and $Z_3$ have a mass splitting of only 1\%, 
$Z_4$ has no other nearby states, $Z_{5,6}$ are nearly degenerate, 
and so forth.  In addition, the pair of states become more degenerate
as the KK mode number increases.
This KK mass spectrum is more easily understood by
examining Figs. \ref{figmasses} and \ref{massspec}, where the
$W$ and $Z$ KK spectra are displayed.  Note that the $W$ KK tower
has a conventional mass spectrum, and each $W$ KK mode coincides
with the pair of degenerate $Z$ KK states.

The couplings of the SM fermions to the massive $Z_n$ KK tower
states can be written in the suggestive form
\be
{g_{Z_n}\over c_w}(T^f_{3L}-s_n^2Q^f)\,,
\ee
with  $c_w=\cos\theta_w^{os}$ and $T^f_{3L}(Q^f)$ being the usual
fermion third-component of weak isospin (electric charge).
Matching with the form of the covariant derivative, the parameter $s_n^2$
is found to be given by
\be
s_n^2={-\lambda\chi_n^B(R)\over \chi_n^L(R)-\lambda\chi_n^B(R)}\,,
\ee
with $s^2_1\equiv\sin^2\theta_{eff}$, \ie, the value of the
weak mixing angle obtained on the $Z$-pole.  The values for $s_n^2$
vary significantly, even in sign, as the
KK mode number varies.  For example, for $\kappa=3$,
$s^2_2=0.743\,, s^2_3=-0.109\,,$ and $s^2_4=0.218$.  We note that for
the KK levels which are non-degenerate, the value of $s^2_n$ is
not too far from the on-shell value, $\sin^2\theta_w^{os}\simeq
0.22210$, as defined above.  This can be understood as being due to
the fact that the double states are mixtures of the $\gamma$ and
$Z$ excitations, while the single states are almost pure $Z$ excitations.
Turning to $g_{Z_n}$, we know that in the SM $g_{Z}=g_W$, \ie,
$g_{Z_1}=g_{W_1}$ and it is traditional to define an effective
$\rho$ parameter \cite{rhoeff}
\be
\rho^Z_{eff}={g_{Z_1}^2\over g^2_{W_1}}={g_Z^2\over g_W^2}\,,
\ee
which can be directly calculated once the $g_{Z_n}$ are known.
Matching with the covariant derivative, we find that these
couplings can be written as
\be
{g^2_{Z_n}\over c_w^2}
=\Omega^2g^2_{5L}{ |\chi^L_n(R)-\lambda\chi_n^B(R)|^2\over N_{Z_n}}\,,
\ee
where $\Omega$ is determined numerically as discussed above.
The normalization in the absence of brane kinetic terms is easily
obtained,
\be
N_{Z_n}^0=2\int_R^{R'}dz\,{R\over z}\,\left\{
|\chi_n^L(z)|^2+|\chi_n^R(z)|^2+|\chi_n^B(z)|^2\right\}\,,
\ee
with the factor of 2 being related to the interval of integration
as described above.  In order to determine $N_{Z_n}$ in the more
general case we must return to the two actions in Eqns. (3)
and (4).  To simplify the discussion, we first rescale
each gauge field by its appropriate 5-d coupling, $A_i\to g_{5i}A_i$,
and concentrate solely on the action integrands which we can combine
and write symbolically as
\be
-{1\over 4}F_L^2-{1\over 4}F_R^2-{1\over 4}F_B^2-{1\over 4}F_C^2
-{1\over 4}(c_LF_L^2+c_YF_Y^2+c_sF_C^2)r_c\delta(y)\,.
\ee
From this it is clear how to normalize fields \cite{dhrbt} which are purely
composed of $A_L$ or $A_C$ as in the case of the $W^\pm$ above.
The difficulty with the remaining fields is that
both the gauge fields and brane terms are a mixture of the two
bulk fields as can be seen from the definition of $Y$ in 
Eq. (16).  Rewriting the $Y$ fields in terms of
$A_R^3$ and $B$, substituting the KK decomposition into the
respective field strength tensors for the neutral fields, and
neglecting the QCD terms, we see that symbolically
\bea
& & -{1\over 4}F_L^2-{1\over 4}F_R^2-{1\over 4}F_B^2
-{1\over 4}(c_LF_L^2+c_YF_Y^2)r_c\delta(y)\to \nonumber\\
& & |\chi_L|^2+|\chi_R|^2+|\chi_B|^2+c_Lr_c|\chi_L|^2
\delta(y)+c_Yr_c \left| { g_{5R}\chi_B+g_{5B}\chi_R\over
\sqrt{g^2_{5R}+g^2_{5B}}} \right|^2\delta(y)\\
& & = |\chi_L|^2(1+c_Lr_c\delta(y))+|\chi_R|^2+|\chi_B|^2
+c_Yr_c\left| {\kappa\chi_B+\lambda\chi_R\over\sqrt{\kappa^2
+\lambda^2}}\right|^2\delta(y)\,.\nonumber
\eea
Allowing for the extension of the integration range, this gives
\bea
N_{Z_n} & = & \int_R^{R'}dz\, {R\over z}\, \left\{
|\chi_L^n(z)|^2(2+c_Lr_c\delta(z-R))+2|\chi_R^n(z)|^2
+2|\chi^n_B(z)|^2\right.\nonumber\\
& + & \left. c_Yr_c{|\kappa\chi^n_B(z)+\lambda\chi^n_R(z)
|^2\over\kappa^2+\lambda^2}\delta(z-R)\right\}\,,
\eea
which reduces to the result above when the $c_i$ are neglected.
Note that this expression also tells us how to normalize the
photon field, which is a constant (\ie, $z$-independent) with the
substitutions  $\chi_L^\gamma=\alpha_L\,, \chi_R^\gamma=
\alpha_L/\kappa\,,$ and $\chi_B^\gamma=\alpha_L/\lambda$.  We
will return to this point below.  Given $N_{Z_n}$, the
$g^2_{Z_n}$ are calculable and $\rho^Z_{eff}$ can be directly
determined; we find that in all cases $|\rho^Z_{eff}-1|\lsim 10^{-4}$.
As in the case
of the $W$ KK tower, these couplings are observed to decrease rapidly as the
KK mode number increases.  For example, if $\kappa=3$, the first
KK excitation above the $Z$ has a coupling strength which is only 
$\sim 11\%$ of the SM $Z$ boson.

Returning to the case of the photon, we note from the form of
the covariant derivative that it couples as
\bea
& & g_{5L}T^f_{3L}\chi_L^\gamma+g_{5R}T^f_{3R}\chi_R^\gamma
+g_{5B}{B-L\over 2}\chi_B^\gamma\nonumber\\
& & = g_{5L}\alpha_L\left(T^f_{3L}+T^f_{3R}+{B-L\over 2}\right)
\equiv g_{5L}\alpha_LQ^f\,,
\eea
apart from a normalization factor which can be determined
directly from $N_{Z_n}$ above, giving
\be
N_\gamma=2\pi r_c\alpha_L^2\left( {\kappa^2+\lambda^2+\kappa^2
\lambda^2\over\kappa^2\lambda^2}\right)\left\{ 1+{1\over\pi kr_c}
{\kappa^2\lambda^2\delta_L+(\kappa^2+\lambda^2)\delta_Y\over
\kappa^2+\lambda^2+\kappa^2\lambda^2}\right\}\,.
\ee
We thus obtain the $\alpha_L$ independent quantity
\be
e^2\equiv {g_{5L}^2\alpha_L^2\over N_\gamma}\,,
\ee
from which we can define the mixing angle
\be
\sin^2\theta_{eg}\equiv {e^2\over g^2_{W_1}}\,,
\ee
where $g^2_{W_1}$ has been previously defined.

\begin{figure}[htbp]
\centerline{
\includegraphics[width=9cm,angle=90]{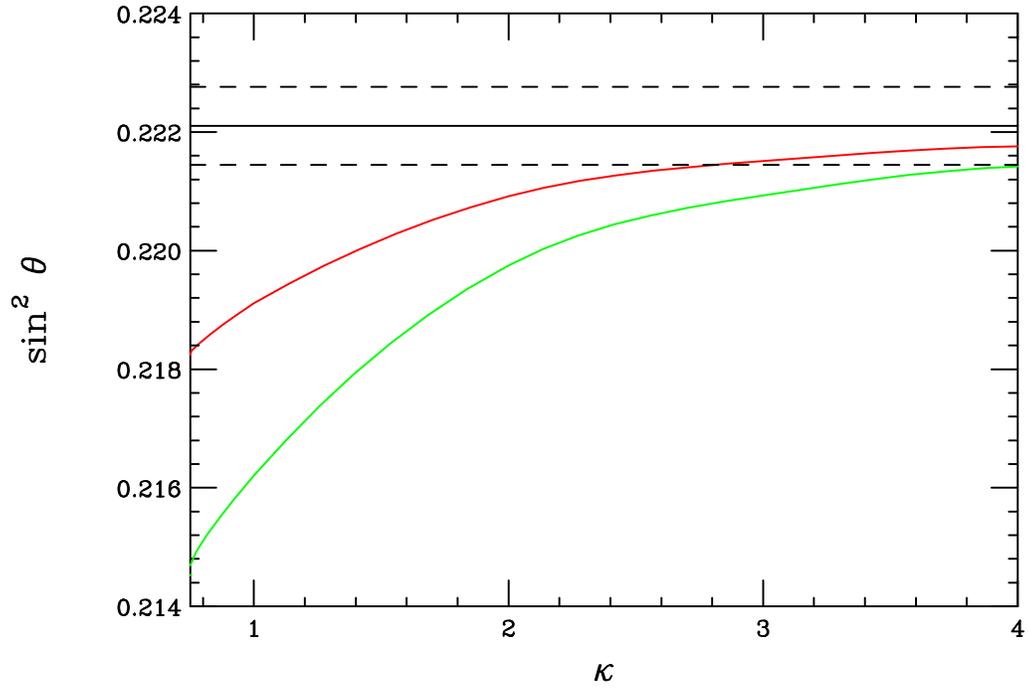}}
\vspace*{0.1cm}
\caption{$\sin^2\theta$ in each of the three definitions as
a function of $\kappa$.  The black (upper), red (middle), and green
(lower) curves correspond to the schemes $\sin^2\theta_w^{os}$,
 $\sin^2\theta_{eff}$, and $\sin^2\theta_{eg}$ defined in the
text.  The dotted curves show the present
$1\sigma$ errors on $sin^2 \theta_w^{os}$ from measurements of the $Z$ and
$W$ masses.}
\label{sinth}
\end{figure}

Note that in the above discussion we have introduced three different
definitions of the weak mixing angle: ($i$) the on-shell value
$\sin^2\theta_w^{os}$, ($ii$) the effective value on the
$Z$-pole, $\sin^2\theta_{eff}$, and ($iii$) $\sin^2\theta_{eg}$.
In the SM, {\it at tree-level}, the values of
these three definitions are, of course, equivalent.  In
the WHM, they need not be in general; however, if the model is to
be consistent with experiment, it is clear that these three
quantities must be reasonably close numerically.  Fig. \ref{sinth}
shows these three definitions of $\sin^2\theta_w$ as functions
of the parameter $\kappa$, where $\sin^2\theta_w^{os}$ is, of
course, $\kappa$ independent.  As a rough guide, the figure 
also shows the current $1\sigma$ errors on the value of
$\sin^2 \theta_w^{os}$ arising from the measured $W$ and $Z$ mass 
uncertainties.  From this figure, we see that as
$\kappa$ increases the three values of $\sin^2\theta_w$ merge
together.  This is due to the KK masses becoming heavier as well
as the strengthening of the SU(2)$_R$ couplings
associated with the custodial symmetry which forces the WHM
to become more like the SM.  Clearly, for values of $\kappa
\simeq 3-4$, the three definitions of the weak mixing angle
are quite close numerically.  It is interesting to note that this
model predicts $\sin^2\theta_{eff}$ to be somewhat smaller
than the on-shell value, \eg, for $\kappa=3$,
$\sin^2\theta_w^{os}-\sin^2\theta_{eff}\simeq 0.0006$.  This
slightly lower value of $\sin^2\theta_{eff}$ is suggestive
of the low value obtained from LEP and SLD \cite{lepewwg}
from measurements of the leptonic couplings of the SM $Z$.

Before we further discuss the electroweak parameters in the next section,
we will conclude this section by examining the KK tower
associated with the gluon.  In analogy to the case of the 
photon, the massless gluon zero-mode has a flat,
$z$-independent wavefunction.  This implies that the conventional
strong coupling can be defined directly via the zero-mode coupling to
fermions following from the boundary conditions and
the KK decomposition.  We thus can write
\be
g_s^2={g_{5s}^2\over 2\pi r_cZ_0}\,,
\ee
where
\be
Z_0=1+{c_s\over 2\pi}=1+{\delta_s\over\pi kr_c}\,.
\ee
Note that to maintain $Z_0>0$, $\delta_s\geq -\pi kr_c$ is required.
Solving for $g^2_{5s}$ we obtain
\be
g^2_{5s}={2\pi r_c g_s^2\over 1-g_s^2/\tilde g^2_s}\,,
\ee
so that
\be
\delta_s=\pi kr_c {g_s^2\over\tilde g_s^2 - g_s^2}\,.
\ee
Since $1/\tilde g_s^2=1/\tilde g_L^2\cdot(\beta_s/\beta_L)$
is known, $\delta_s$ can be directly calculated.  Taking
$\alpha_s=0.118$ we obtain
\be
\delta_s\simeq -29.14\,,
\ee
independent of $\kappa$.

Knowing the value of $\delta_s$, we can now determine the
gluon KK spectrum.  Note that this value of $\delta_s$ is
not far away from the critical region of $\delta=
-\pi kr_c\simeq -35.4$ discussed above, where the KK spectrum
and couplings become highly perturbed as shown in our earlier
work \cite{dhrbt}.  In fact, for $\delta_s\leq -\pi kr_c$, the
system becomes unphysical as ghost states appear.  For the
value of $\delta_s$ computed above, the first gluon KK excitation,
$g_1$, is pushed upwards in mass by $\simeq 10\%$ in comparison
to what would be naively expected for smaller values of the
brane term, and hence $m_{g_1}$ is roughly 200 GeV heavier than the
first gauge KK excitation.  The mass splitting for the 
higher gluon KK states are
similar to those of the $W$ KK tower.  

\begin{figure}[htbp]
\centerline{
\includegraphics[width=9cm,angle=90]{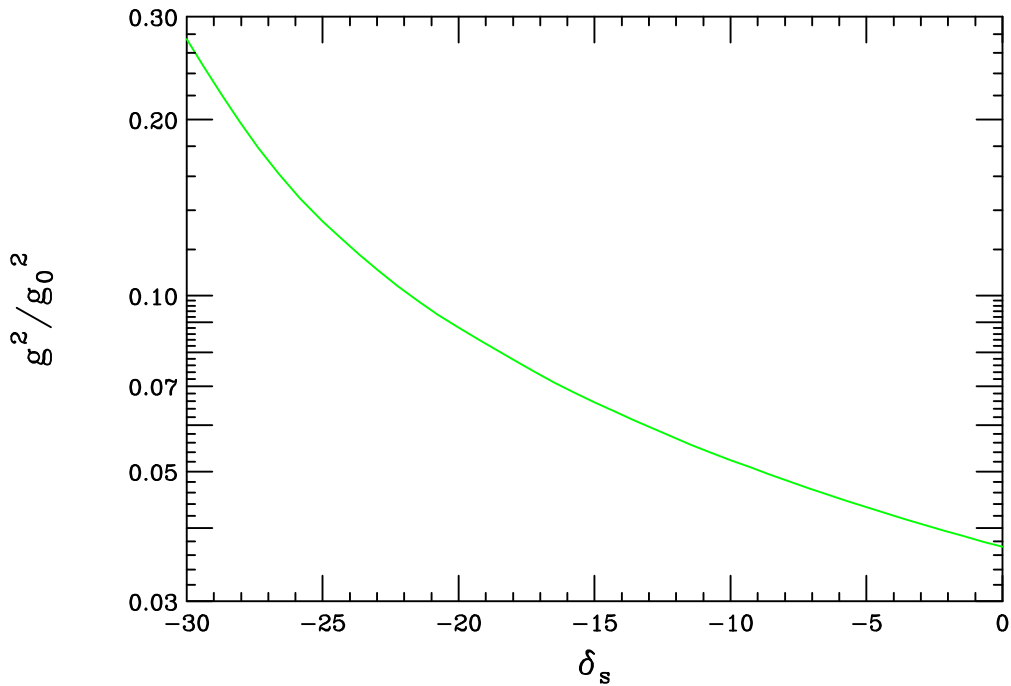}}
\vspace*{0.1cm}
\caption{Behavior of the coupling of the SM quarks to the first gluon
KK excitation as a function of the brane term $\delta_s$, demonstrating
the rapid growth in the coupling as $\delta_s\to -\pi kr_c$.}
\label{glue}
\end{figure}

The couplings of the gluon KK tower states can be directly
calculated as in the $W$ and $Z$ cases above from the covariant
derivative,
\be
{g^2_{s_n}\over g_s^2} = 2\pi r_cZ_0{|\chi_n^C(R)|^2\over
N_{s_n}}\,,
\label{gluecoup}
\ee
where
\be
N_{s_n}=\int_R^{R'}\, dz\, {R\over z} |\chi_n^C(z)|^2[2+c_sr_c
\delta(z-R)]\,.
\ee
Here, we note that $g^2_{s_0}=g^2_s$, the usual QCD
coupling.  The
coupling of the first gluon KK state is displayed as a function
of the strong brane term in Fig. \ref{glue}, where we see that
the KK states of the gluon are more strongly coupled, scaled to
the zero-mode coupling strength, as compared to the other KK
towers.  For the higher KK levels, the ratio $g_{s_n}^2/g_s^2$ does not
decrease as quickly as in, \eg, the case of the corresponding
$W$ boson KK tower couplings.  For example, $g^2_{s_2}/g_s^2\simeq
0.233$, while we previously found  $g^2_{W_2}/g^2_W\simeq 0.043$.
This is due to the large magnitude of $\delta_s$.

\section{Electroweak Oblique Parameters}

As discussed in the previous section, the WHM leads to a complete 
determination of the couplings of the $W\,,Z$ and gluon (as well 
as their KK towers) as a function of $\kappa$.
We showed that the gauge boson zero-mode couplings to fermions 
are slightly different from their corresponding tree-level values 
in the SM, resulting in shifts from the SM 
expectations for the precision electroweak observables.  
It has become common practice in the literature to describe the influence
of many classes of new physics on electroweak precision data at the one loop
level through the use of the oblique parameters $S\,,T\,,U$ {\cite{PT}}. In the
present model we have observed substantial deviations from SM expectations, 
\eg, the three distinct values of $sin^2\theta$, already at the tree level.
Similar parameter shifts are known to exist in the case of other sources of
new physics, such as in the case of a simple $Z'$ model {\cite{stuzp}}. Though
such corrections are not oblique, it has been shown{\cite {stuzp}} that the
shifts in several electroweak observables can be parameterized in a manner
similar to that of $S\,,T\,,U$. 
In order to not  confuse any such parameterization
with the usual oblique parameters 
$S\,,T\,,U,$, we will denote these pseudo-oblique parameters as $\Delta S\,,
\Delta T$\,, and $\Delta U$. We wish to emphasize that these quantities are
being introduced solely as a device to demonstrate the deviations of the
present scenario from SM expectations and they are not to be interpreted as
the ordinary oblique corrections.

\begin{figure}[htbp]
\centerline{
\includegraphics[width=9cm,angle=90]{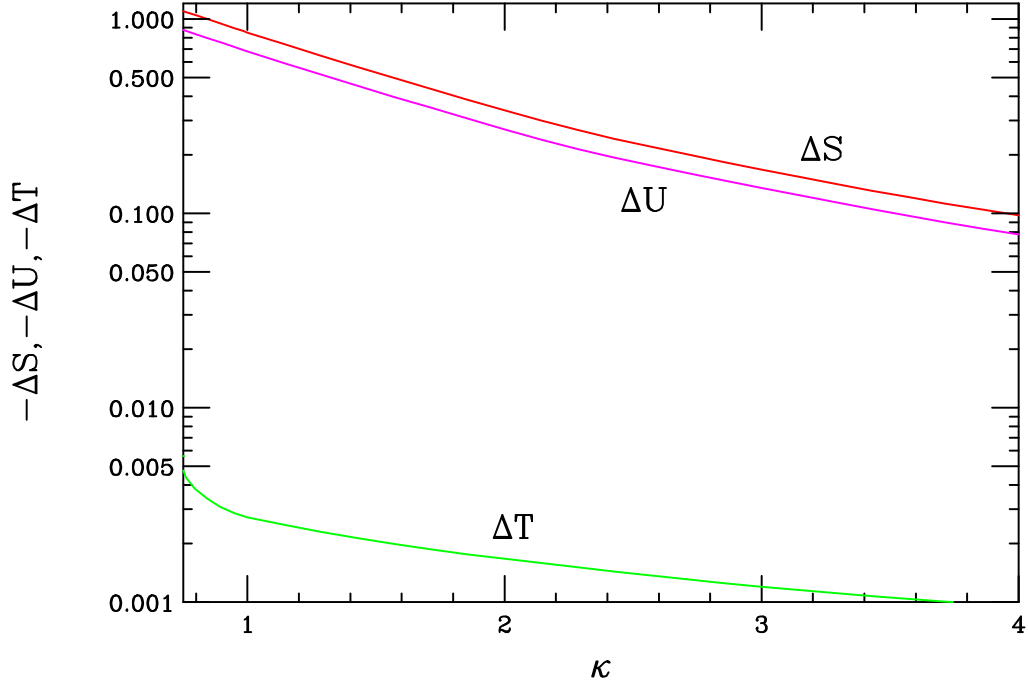}}
\vspace*{0.1cm}
\caption{Shifts in the values of the pseudo-oblique parameters $\Delta S$, 
$\Delta T$, and $\Delta U$
as a function of $\kappa$ from the tree-level analysis discussed in
the text.}
\label{splot}
\end{figure}

We take $\alpha$, $M_Z$, and $G_F$ to be input parameters in
performing our fit to the electroweak measurements.  Usually, when
fitting the electroweak data the most important
set of quantities to examine  is $M_W$, $sin^2 \theta_{eff}$
and the width for either $Z\to l^+l^-$ or $\nu \bar \nu$, \ie, the invisible
$Z$ width. (Here we will employ the invisible width.)  These quantities
are either very precisely measured
or are most unambiguously sensitive to new physics.
We can then {\it parameterize} any deviations of these quantities away from
their SM expectations through the usual definitions, employing
the pseudo-oblique parameters $\Delta S$ \etc.   \cite{PT,obstu}
\bea
\sin^2\theta_{eff} & = & \sin^2\theta_0 + {\alpha\Delta S\over
4(c_w^2-s_w^2)} - {c_w^2s_w^2\alpha\Delta T\over c_w^2-s_w^2}
\,,\nonumber\\
M_W^2 & = & M^2_{W_{SM}} \left[ 1 - {\alpha\Delta S\over 2
(c^2_w - s_w^2)} + {c_w^2\alpha\Delta T\over c_w^2-s_w^2}
+{\alpha\Delta U\over 4 s_w^2}\right]\,,\\
\Gamma_\nu & = & \Gamma_{\nu_{SM}}(1+\alpha\Delta T)\,,\nonumber
\eea
where $\alpha$ is the fine-structure constant.  Note that here, we
are simply exchanging the shifts from the SM predictions for the observables
on the left-hand side of this equation for the pseudo-oblique parameters. 
We write $\Delta S(T,U)$
as shifts in these parameters away from their exact value at 
tree-level in the SM, \ie, we recover the SM when the pseudo-oblique
parameters vanish. Since we are comparing with the SM at 
tree-level we have $\sin^2\theta_0=\sin^2\theta_w^{os}$.  The ratio
$\Gamma_{\nu}/\Gamma_{\nu_{SM}}$ is equal to $\rho^Z_{eff}$,
using the notation of the previous section.  Lastly, we have again
imposed the requirement that
$M_W$ be in agreement with its SM value as defined
by experiment, so that the expression in brackets on the 
right-hand side of the equation must vanish, thus forcing a 
relationship between the pseudo-oblique parameters.  Since for all values
of $\kappa$, $\rho^Z_{eff}$
is found to differ from unity only at the order of a few $\times 10^{-5}$,
it is clear that $\Delta T$ is very small.  The expression
for $M_W$ then yields
\be
\Delta U \simeq {2s_w^2\over c_w^2-s_w^2}\, \Delta S\,.
\ee
Using the values of $\rho^Z_{eff}$ computed above and 
$\sin^2\theta_{eff}$ from the previous section, we can determine
the pseudo-oblique parameters as a function of $\kappa$.  This
is displayed in Fig. \ref{splot}.  Here, we see that $\Delta T$ is
very small as expected, $\Delta U$ tracks $\Delta S$, and $\Delta 
S$ falls rapidly in magnitude as $\kappa$ increases, as expected.  
The main point of this figure is to demonstrate that the pseudo-oblique 
parameters fall
rapidly to zero as the value of $\kappa$ increases; this is as expected 
since this limit approaches the SM.

The most recent fit to the oblique parameters has been performed by 
Erler {\cite {jens}} using the data presented at the 2003 summer 
conferences \cite{lepewwg}. 
The results for this fit are highly correlated; using $m_H=117$ GeV, Erler 
obtains the $1\sigma$ constraints $S=-0.13\pm 0.10$, $T=-0.17\pm0.12$ and 
$U=0.22\pm 0.13$. Slightly negative values of $S,T$ are favored while the fit 
prefers slightly positive values of $U$.   
While we cannot directly compare to the data it
is clear that the results of Fig \ref{splot}, as
well as Fig. 4, strongly suggest that a reasonably large value of
$\kappa \gsim 3$, approximately reproduces the SM at tree-level.

We recall that at loop level,  $S$, $T$, and $U$ 
are traditionally determined from the gauge 
boson self-energies \cite{PT}.  Note that we are now working with the
formal definitions of $S\,, T$ and $U$ from \cite{PT}.
Loop contributions are of order $\alpha$, so they may also be 
important compared to the tree-level values discussed above.
To leading order in $1/(\pi kr_c)$ the 
wavefunctions of the $W^{\pm}_1$ 
and $Z_1$ are flat in $z$, so we can calculate loop contributions in this 
approximation; the corrections will be of order $\alpha/(k\pi r_c)
\simeq 2\times 10^{-4}$, and can be safely ignored.  For the photon, 
of course, the wavefunction is flat to all orders. The coupling of an 
approximately flat zero-mode to two excited modes is then given by, \eg,
\begin{gather}
g_{5L}\Omega \int_{R}^{R'} dz\, \frac{R}{z}\
\frac{\chi_1^{Z}\chi_n^{W}\chi_m^{W}[2+c_Lr_c\delta(z-R)]}
{(N_{Z_1} N_{W_n} N_{W_m})^{1/2}}= g_{5L} \delta_{nm}
\end{gather}
by the orthonormality condition in Ref. \cite{dhrbt}. 
This means that the couplings of the KK $W$ 
modes (which participate in the loop of the $\gamma/Z$ self-energy
diagram) to the exterior $Z$ or $\gamma$ are exactly 
the same as the SM triple gauge 
couplings in this limit. In particular, the coupling of an excited $W$ to 
the hypercharge boson is zero. We can write $S$ as \cite{Luty:1992fe}
\begin{gather}
S = 
-16\pi\left.\frac{\partial \Pi_{ZY}(q^2)}{\partial q^2}\right|_{q^2=0},
\end{gather}
where $\Pi_{ZY}(q^2)$ is the self-energy mixing between the $Z$ 
and the hypercharge boson (in this example) 
through $W$ loops. So we conclude that, 
at order $\alpha$, $\Delta S= 0$. We also expect the 
contribution to $T$ to be small due to the presence of the custodial
symmetry and because the mass 
splittings between the excited $W$ and $Z$ bosons are small.
Hence the KK loop contributions to the oblique parameters can be
safely ignored.

A more serious problem arises from the fact that the Higgs boson is no
longer in the spectrum, and hence cannot run in loops. To estimate
the one loop values of $S$, $T$, and $U$ correctly, one would need a
procedure for systematically removing the effects of the Higgs loops from
the precision electroweak observables. It is not clear how this can be
accomplished easily, due to the non-gauge invariant nature of the
relevant graphs.

It is also possible that there are higher dimension operators 
localized on the $\tev$ brane that violate $S$, since there is no 
symmetry to prevent them ($T$ is protected by the custodial SU(2)). 
The size of these operators will naively be 
$M_Z^2/\Lambda_\pi^2 \approx 10^{-4}$, leading to contributions to $S$ 
of order $\frac{1}{\alpha} M_Z^2/\Lambda_\pi^2 \approx 10^{-2}$.

In Ref. \cite{Barbi}, the precision electroweak constraints on the 
WHM model, within a less general
parameter space, were considered.  Qualitatively, we agree with their
conclusion that in the regime where many KK modes lie below the IR
cutoff scale $\sim \Lambda_\pi$ of the warped space (corresponding to the
regime of weakly interacting distinct states), the WHM is excluded
by precision electroweak data.  In our approach, a similar conflict arises
between the electroweak data and unitarity, where the former requires the
absolute scale of higher KK modes to lie above $\sim 2$ TeV, and the
latter demands the opposite.

\section{Perturbative Unitarity in Gauge Boson Scattering}

An important function of the Higgs boson in the SM is to
insure the perturbative unitarity of the broken gauge theory. 
In this Higgsless model,
we would like to test the claim in Ref. \cite{CsakiI,CsakiII} 
that the KK modes will be able to insure
the unitarity in place of the Higgs.

\begin{figure}[t]
\centerline{
\includegraphics[bb = 149 364 562 523,width=12cm]{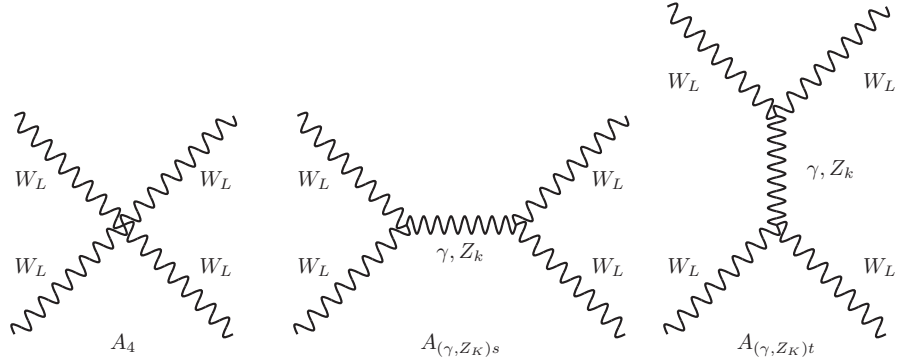}}
\caption{
Feynman diagrams for the tree-level amplitudes contributing to $W_L^+ W_L^-$
scattering.}
\label{wwgraphs}
\end{figure}

The classic test of perturbative unitarity is the elastic scattering of two
longitudinally polarized gauge
bosons, $W_L^+ W_L^- \to W_L^+ W_L^-$ \cite{lqth}. 
This amplitude receives tree-level
contributions from the four-$W$ vertex, and from the three-boson vertices
through exchange of a single neutral gauge boson in the $s$- and $t$-channels, 
as shown in Fig. \ref{wwgraphs}. The diagram involving the four-boson 
vertex contains terms that grow like $s^2$, $s$ and $s^0$, as well
as innocuous terms involving powers of $1/s$. For the scattering to
respect unitarity, the terms that grow with $s$ must cancel against 
those arising from other graphs in the theory.
Csaki \etal\ \cite{CsakiI} 
have investigated the behavior of these terms at large
$s$. Strictly speaking, the expansion they 
performed is only valid at energies `above' all the
KK masses. In practice, however, it is a good approximation to take
values of $\sqrt s$ above a sufficiently
large number of KK modes. In that region, as shown in Ref.
\cite{CsakiI}, there are two necessary conditions for
the terms which grow with energy in the 4-point contribution
 to be cancelled by those from the one-boson exchange graphs:
\begin{align}
g^2_{nnnn} & = \sum_{k}g^2_{nnk}\,,\nonumber\\
4 g^2_{nnnn} M_n^2 & = 3\sum_{k}g^2_{nnk}M^2_k\,.\label{eq:sumrule2}
\end{align}
Here $g^2_{nnnn}$ is the coupling of four gauge bosons with KK-number $n$,
and $g_{nnk}$ is the three boson coupling between two states with
KK-number $n$ and one with KK-number $k$.
The first of these conditions insures the cancelling of terms in the
amplitude that grow like $s^2$, and is guaranteed by the original gauge
invariance. The second condition is required for the cancellation of
the terms that grow like $s$, and it is not
trivial that it will be satisfied in the present model.  For the
case of ordinary $W$ scattering, $n=1$.

To test these conditions we have examined numerically the case of
$W^+_LW^-_L$ scattering, since this is an important
process, and will be measured 
at future colliders. The relevant couplings are given by
\bea
g^2_{1111} & = & g^2_{5L}\Omega^2
\int_R^{R'}dz\, {R\over z}\, {1\over N_{W_1}^2} 
\left(|\chi^{L^\pm}_{1}|^4 [2+c_Lr_c\delta(z-R)] 
+ 2\kappa^2|\chi^{R^\pm}_{1}|^4\right)\,,\\
g_{11k} & = & g_{5L}\Omega 
\int_R^{R'}dz\, {R\over z}\, {1\over N_{W_1} \sqrt {N_{Z_k}}} 
\left( |\chi^{L^\pm}_{1}|^2\chi^{L^0}_{k}[2+c_Lr_c\delta(z-R)]
+2\kappa |\chi^{R^\pm}_{1}|^2\chi^{R^0}_{k} \right)\,,\nonumber
\eea
where $N_{W_1}$ and $N_{Z_k}$ are the normalization factors given 
above. For the first sum rule we also need the coupling of two
$W^\pm_1$ bosons to the photon, which is just $e$ by gauge invariance.

\begin{centering}
\begin{figure}
\includegraphics[width=12cm,angle=0]{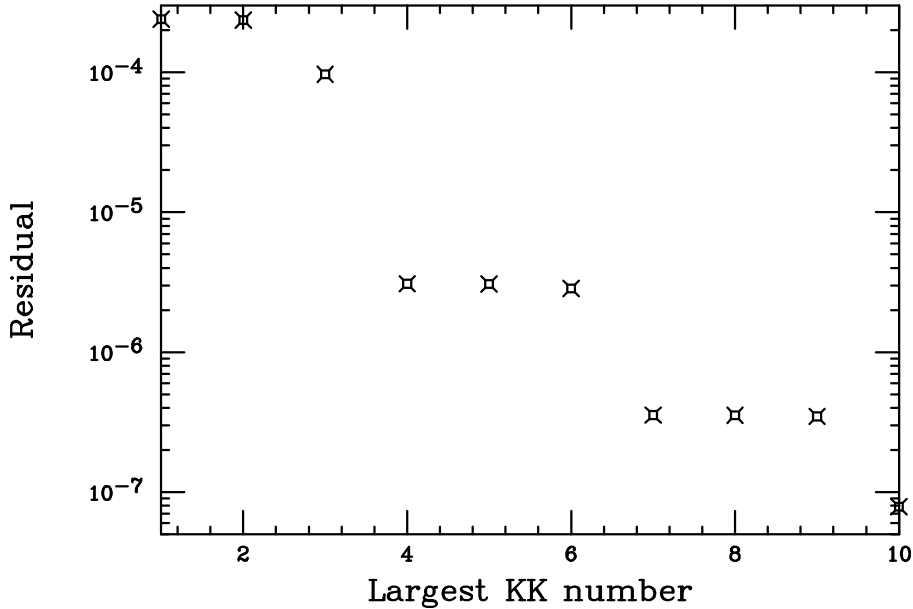}
\caption{The residual of the sum rule $1 = \sum_k g^2_{11k}/g^2_{1111}$ as a
function of the highest KK state included in the sum. This shows that the
sum rule is converging, and hence the cross section will behave like $1/s$
at asymptotically large $\sqrt{s}$.  Here we have assumed
that $\kappa=3$ for purposes of demonstration.}
\label{fig:convergence}
\end{figure}
\end{centering}

We have numerically evaluated $g^2_{1111}$ and $g_{11k}$ for $k$ extending
over the photon, the $Z_{1}$, and the first 9 (or more) higher excited 
states for the entire range of $\kappa$. The
agreement with the sum rules
is quite good, and was observed to rapidly improve as more states
were added. If, \eg, $\kappa=3$, the
residuals of these sum rules after including the first
9 excited states are
\begin{align}
1 - \sum_{k=\gamma,1}^{10}
\frac{g^2_{11k}}{g^2_{1111}}& = 7.85\times 10^{-8}\,,\nonumber\\
1 -
\frac{3}{4}\sum_{k=1}^{10}\frac{g^2_{11k}}{g^2_{1111}}\frac{M^2_{Z_k}}
{M_{W_1}^2}& = 1.96\times 10^{-3}.\label{eq:residual2}
\end{align}
This shows that the sum rules are being satisfied, and so in the
asymptotic region the cross section will indeed fall like $1/s$.
The convergence of these sums as more KK states are added can be seen in Fig.
\ref{fig:convergence}.

The sum rules, however, are necessary, but not sufficient conditions for
perturbative unitarity. 
In particular, the amplitude for $W_L^+W_L^-$ scattering, which
grows like $s^2$ near a few times 
$M_W^2$, could grow too large before sufficiently
many KK modes are passed. There is also a term formally independent of 
$s$, the
coefficient of which could grow as more and more KK modes are included. It
is possible that this term will also contribute to unitarity
violations. To investigate this issue we examine the full amplitude for
$W_L^+ W_L^-$ scattering.

The amplitudes due to photon and $Z$ exchange have been previously
computed. Simple
modifications of the formulae in Ref. \cite{Duncan:vj} gives
\begin{align}
A_{s\gamma} & = -\frac{1}{16}ie^2s^2\beta^2 (3-\beta^2)^2 
\cos\theta\,,\nonumber\\
A_{sZ_k} & = -\frac{1}{16}i g^2_{11k}
\frac{s^3}{s-\xi_{Z_k}}\beta^2(3-\beta^2)^2\cos\theta\,,\nonumber\\
A_{t\gamma} & = -\frac{ie^2 s^3}{32t}\left[
\beta^2(4-2\beta^2+\beta^4)+\beta^2(4-10\beta^2+\beta^4)\cos\theta
\right.\notag\\
&\left.+(2-11\beta^2+10\beta^4)\cos^2\theta +\beta^2\cos^3\theta\right]
\,,\\
A_{tZ_k} & = -\frac{ig_{11k}^2 s^3}{32(t-\xi_{Z_k})}\left[
\beta^2(4-2\beta^2+\beta^4)+\beta^2(4-10\beta^2+\beta^4)\cos\theta
\right.\notag\\
&\left.+(2-11\beta^2+10\beta^4)\cos^2\theta +
\beta^2\cos^3\theta\right]\,,\nonumber\\
A_4 & = -\frac{1}{16}i g^2_{1111} s^2(1+2\beta^2-6\beta^2\cos\theta-
\cos^2\theta)\,,\nonumber
\end{align}
where $\xi_{Z_k} = M_{k}^2/M_W^2$, $t = -\frac{1}{2}s\beta^2(1-\cos\theta)$,
and $\beta = \sqrt{1-4/s}$, and the labels refer to $s$ and $t$-channel exchanges. Here $s$ and $t$ have been scaled to $M_W^2$.
As is well known, in the SM the sum of these amplitudes 
grows like $s$. This growth is cancelled by
the Higgs contributions
\begin{align}
A_{sH} &= -\frac{1}{16}ig^2s^2(1+\beta^2)^2\frac{1}{s-\xi_H}\,,\nonumber\\
A_{tH} &= -\frac{1}{16}ig^2s^2(\beta^2-\cos\theta)^2\frac{1}{t-\xi_H}\,,
\end{align}
with $\xi_H = m_H^2/m_W^2$.

\begin{centering}
\begin{figure}
\includegraphics[width=12cm,angle=0]{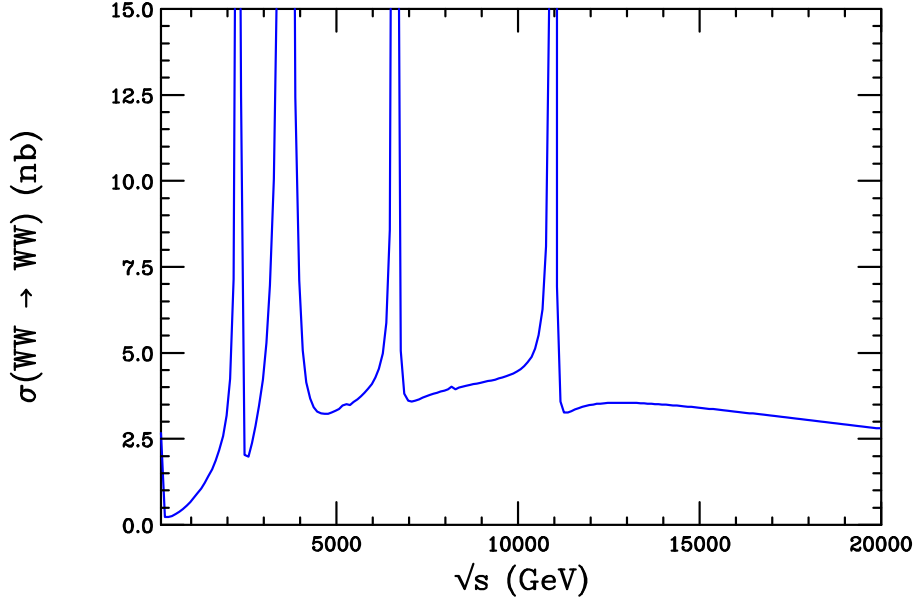}
\caption{The cross section 
for $W_L^+ W_L^- \to W_L^+ W_L^-$ scattering with the first 10
KK states included. A heavy fake state has also 
been included with a mass of $14.7\
\tev$ and coupling $g = 2.8\times 10^{-4} g_{1111}$ to complete the sum
rules and show that the cross section falls like $1/s$ asymptotically. No
attempt has been made to smooth the poles at the KK resonances.
Here we have assumed that $\kappa=3$ for purposes of demonstration and
have set $z_0=0.98$.}
\label{fig:sigma10kk}
\end{figure}
\end{centering}

As we have seen, in the present model, the terms growing with $s$ are
cancelled at large $s$ by the sum over the KK modes. For intermediate
regions of $s$ we investigate the full amplitude
\begin{gather}
A = A_4 + A_{s\gamma} + A_{t\gamma} + \sum_{k=1}^{\infty}
\left(A_{sZ_k} + A_{tZ_k}\right)\,.\label{eq:wwamplitude}
\end{gather}
For reference, the cross section,
with a cut on the scattering angle $|\cos\theta| \le z_0$, is
\begin{gather}
\sigma = \frac{1}{16 \pi s^2\beta^2}\int_{t_-}^{t_+}dt\,|A|^2\,,
\end{gather}
with $t_{\pm} = (2-\frac{1}{2}s)(1\mp z_0)$. This cross section, summed
over the first 10 KK modes, is shown in Fig. \ref{fig:sigma10kk},
taking $z_0=0.98$.  To demonstrate the asymptotic behavior 
for the case $\kappa=3$, we have inserted a heavy
fake state with mass $m_{heavy}=14.7$ TeV and coupling
$g=2.8\times 10^{-4}g_{111}$ chosen to cancel the residuals in
Eq. (68).  This heavy fake state is intended to numerically compensate 
for extending the KK sum out to infinity.
When this state is included, the
cross section is seen to fall like as expected. However, it is clear
that while including 10 KK states is enough to flatten the cross section,
as seen in the region below the fake state, it is not enough to make it
fall with $s$.

\begin{centering}
\begin{figure}
\includegraphics[width=12cm,angle=0]{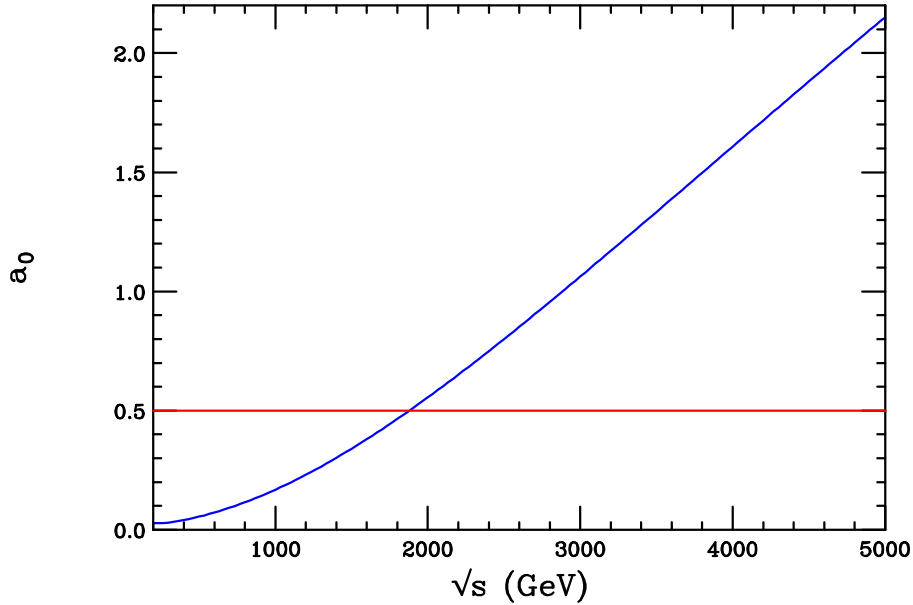}
\caption{The real part of the zeroth partial wave amplitude for 
$W_L^+ W_L^- \to W_L^+W_L^-$ scattering as a function of $\sqrt{s}$. 
The first 10 KK states have been
included. We have taken $\kappa=3$ and $z_0=0.98$.
Unitarity is violated if this amplitude exceeds $1/2$, which is
seen to occur at $\sqrt{s} \approx 2\ \tev$.}
\label{fig:partialwave}
\end{figure}
\end{centering}

A good test of the unitarity of this scattering process is that the first
partial wave amplitude should be bounded for all $s$ \cite{lqth}
\begin{gather}
|\Re(a_0)| = \left|\Re\left(\frac{1}{32\pi} \int_{-1}^1 d\cos\theta
 (-iA)\right)\right|\le \frac{1}{2}. 
\end{gather}
We have calculated this quantity for the amplitude in
Eq. (\ref{eq:wwamplitude}), again summing over the first 10 (or more) 
KK modes. Our
result is shown in Fig \ref{fig:partialwave} for the case $\kappa=3$. 
Unitarity is clearly violated at a
center of mass energy of $\sqrt{s} \approx 2.0\ \tev$, below the mass of
the first KK mode. Note that this is only slightly better than the value
obtained for the 4-d Standard Model without a Higgs, where unitarity breaks
down at $\sqrt{s} \approx 1.7\ \tev$. The problem can be traced to the
fact that the first higher excited modes are too heavy to have much
influence before unitarity is violated. So, while the cross section will
behave like $1/s$ at {\it asymptotically} high energies, unitarity is violated
before that regime sets in.

\begin{figure}
\includegraphics[width=12cm,angle=0]{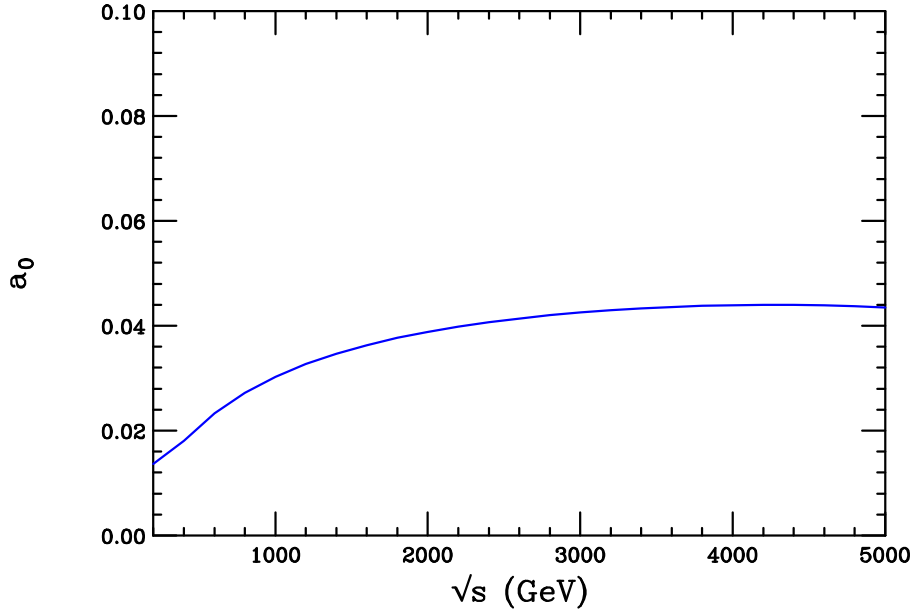}
\caption{The real part of the zeroth 
partial wave amplitude for $W_L^+ W_L^- \to W_L^+
W_L^-$ scattering in the flat space equivalent of the WHM. The first 10 KK
states have been included and we have taken $\kappa=3$ and $z_0=0.98$.
Unitarity is violated if this amplitude exceeds
$1/2$, which is seen not to occur. In this case the sum rules are almost
saturated well before $\sqrt{s}= 1.7\ \tev$, where the SM
without a Higgs boson violates unitarity.}
\label{fig:partialwaveflat}
\end{figure}

For comparison we have performed the same calculation for the equivalent
theory in flat space (with $\kappa = 1$), as presented in Section 6 of Ref.
\cite{CsakiI}.  Our results are 
shown in Fig. \ref{fig:partialwaveflat}. In that case,
the first excited mode sits at $240\ \gev$, and the spacing between
successive modes is $160\ \gev$. When $\sqrt{s}$ has reached a few $\tev$
many KK modes have been passed and both sum rules are nearly saturated, 
so the terms growing with $s$ are nearly cancelled.  We thus see that
the flat space equivalent theory is well-behaved \cite{hjhe}.

\begin{figure}
\includegraphics[width=12cm,angle=0]{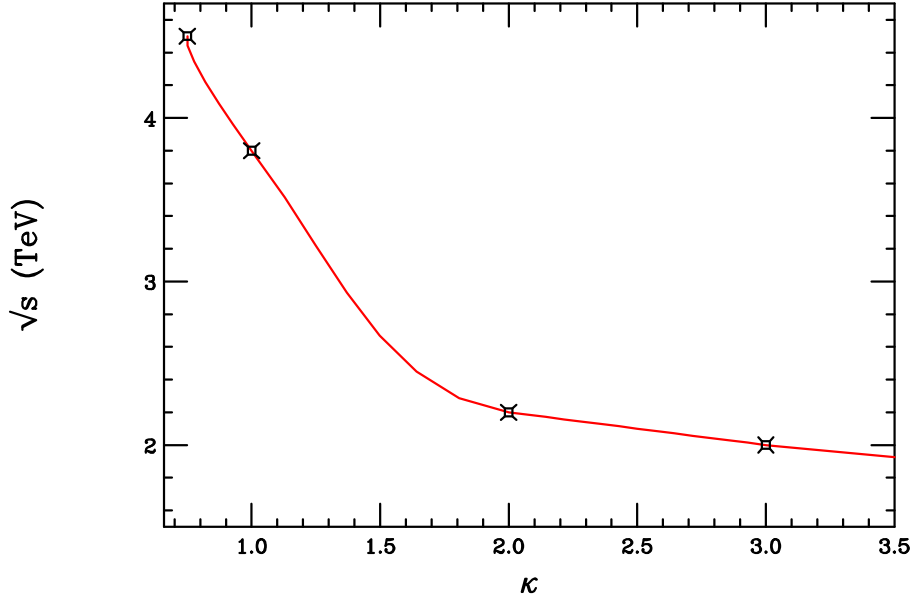}
\caption{The value of $\sqrt s$ at which perturbative unitarity breaks down in
$W_L^+ W_L^- \to W_L^+W_L^-$ scattering as a function of $\kappa$,
taking $z_0=0.98$.  The points
represent the distinct cases for which we numerically computed the
unitarity violation and the curve extrapolates between the points.}
\label{unitkap}
\end{figure}

In order to discern how serious the problem of unitarity violation
is in the WHM, we have repeated the above analysis over the entire 
allowed range of $\kappa$,  even
in the regimes where we do not expect the model to agree with the
precision measurements. In particular, one might expect that if
unitarity is to be respected it would be in the case when the neutral KK states
are as light as possible and where deviations from SM couplings are also
large, \ie, for small values of $\kappa$ \cite{Barbi}.   
However, we find that perturbative
unitarity breaks down for {\it all} values of $\kappa$, as shown in 
Fig. \ref{unitkap}, although the scale where the violation occurs is somewhat
larger when $\kappa$ is small.  We see that for all values of $\kappa$,
perturbative unitarity is violated below any new scale, such as $\Lambda_\pi$
for $k/\mpl\leq 0.1$.
It is possible that the additional brane terms mentioned in Section 4 could
be adjusted to make the theory both unitary and consistent with data.
Whether this can be achieved with or without fine-tuning is not clear.

We also note that the $W_L^+W_L^-$ 
scattering process can proceed by KK graviton
exchange, and that this contribution has the opposite sign in the amplitude,
so one might hope that unitarity could be restored by destructive
interference between the gauge and gravity sectors. However, the ratio of
the KK graviton to gauge exchange amplitudes is roughly
$M_W^2/(g^2\Lambda_\pi^2)\approx 10^{-4}$, 
leading to a strong suppression. Numerically we
find this effect to be insignificant.  A similar situation holds for
the contribution of the radion scalar, which is also present in the
model.  We thus conclude that this model is not a reliable perturbative
framework.

\section{Collider Signals}

Thus far, we have seen that consistency with precision electroweak
data demands that the ratio $\kappa=g_{5R}/g_{5L}$ take on larger
values, such as $\kappa\sim 3$, which strongly enforces the SM limit.  
We have also seen that unitarity
in gauge boson scattering is problematic in the WHM  for such
values of $\kappa$.  Although
the sum rules in Ref. \cite{CsakiI}, which are derived from the amplitude 
for longitudinal
$W$ scattering, are satisfied once enough gauge KK states are included, we 
demonstrated that
the unitarity condition for the zeroth partial wave
amplitude is violated.  
Hence the WHM  is not a weakly coupled model.  
In this section, however, we will take the view that the model may be 
extended or modified,
\eg, with the inclusion of additional non-Higgs states, in such a way as to
restore unitarity.  We thus examine the general collider signatures of 
the gauge and graviton KK states, as these are most likely a
generic feature in any extended Higgsless model based on a warped 
geometry.    

We first examine the signatures of the neutral gauge KK states,
recalling that the mass spectrum of these states and their 
couplings to the SM fields are derived in Section 3.  We will take
$\kappa=3$ throughout this section, in accordance with the constraints
from precision electroweak measurements.  The resulting 
spectrum and fermionic couplings for the first few excited neutral
gauge KK states above the $Z$ are displayed in Table \ref{spect}, where 
the couplings are written in the form
\be
{g_{Z_n}\over c_w}(T^f_{3L}-s^2_nQ^f)\,.
\ee
We see a general trend of decreasing coupling strength,
$g_{Z_n}$, with increasing
KK mode. These couplings are roughly $7-16\%$ (with the
exact value depending on the mode number) of the SM
weak coupling strength, and hence we can expect smaller
production rates for these states.  In addition, we note that the pair of 
nearly degenerate states have different fermionic interactions due
to the parameter $s_n^2$. 
Measuring these couplings would separate the two degenerate states and
uniquely identify this model.

\begin{table}
\centering
\begin{tabular}{|c|c|c|c|} \hline\hline
  & $m_{Z_n}$ (TeV) & $g_{Z_n}/g_{SM}$  & $s^2_n$   \\ \hline
$Z_2$   & 2.30 & 0.106 & 0.743 \\
$Z_3$   & 2.31 & 0.163 & -0.109 \\
$Z_4$   & 3.62 & 0.065 & 0.218 \\ 
$Z_5$   & 5.24 & 0.072 & 0.748 \\
$Z_6$   & 5.26 & 0.113 & -0.104 \\
\hline\hline
\end{tabular}
\caption{Mass spectrum and fermionic couplings for the first five
excited neutral gauge KK states above the $Z$, taking $\kappa=3$.
The KK coupling strength is scaled to the SM weak coupling.}
\label{spect}
\end{table}

\begin{figure}[htbp]
\centerline{
\includegraphics[width=9cm,angle=90]{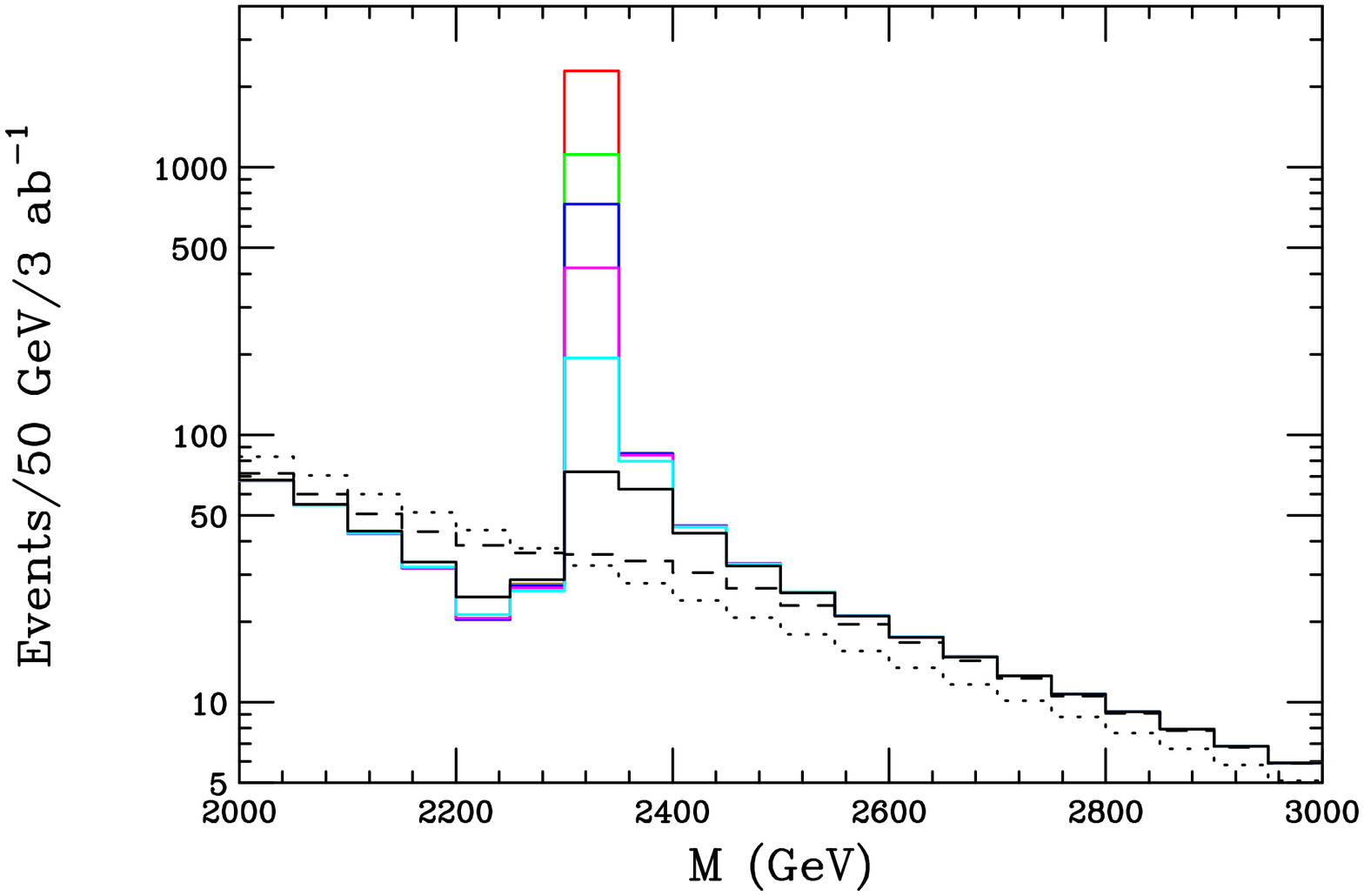}}
\vspace*{0.1cm}
\centerline{
\includegraphics[width=9cm,angle=90]{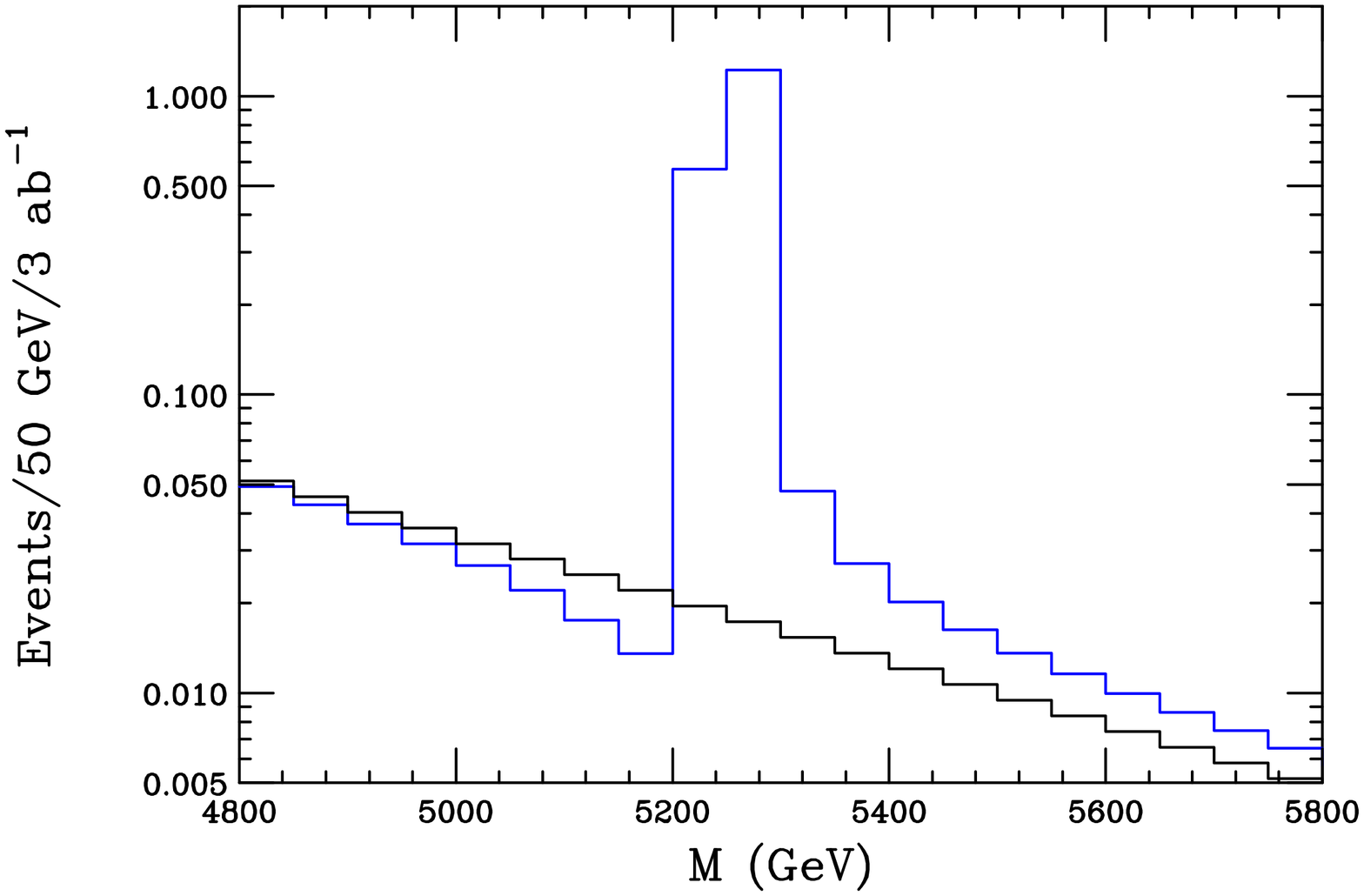}}
\vspace*{0.1cm}
\caption{Top panel: Event rate for Drell-Yan production of the $Z_{2,3}$
gauge KK states, in the electron channel, as a function of the 
invariant mass of the lepton pair at the LHC with 3 ab$^{-1}$ of
integrated luminosity.  The dotted histogram corresponds to the SM
background, while the histograms from the top down (represented by
red, green, blue, magenta, cyan, solid, and dashed) correspond
to letting the width float with a value of $c =1\,,2\,,3\,,5\,,
10\,,25\,,100$.  Bottom Panel: Event rate for Drell-Yan production 
of the $Z_{5,6}$ gauge KK states as a function of the 
invariant mass of the lepton pair at the LHC with 3 ab$^{-1}$ of
integrated luminosity (blue histogram).  The bottom solid histogram 
corresponds to the SM background.}
\label{drelly}
\end{figure}

The classic mechanism for producing heavy neutral gauge bosons in
hadronic collisions is Drell-Yan production, $pp\to Z_n\to\ell^+\ell^-$,
where the $Z_n$ appears as a resonance.
The Drell-Yan lineshape is clearly dependent on the total width of
the $Z_n$, which varies in the WHM
depending on the placement of the fermions.  We have thus allowed for
the total width of the n$^{th}$ gauge KK state to float, 
\be
\Gamma_n = c\,\Gamma_n^0\,,
\ee
where $\Gamma_n^0$ corresponds to the case where all the SM fermions
reside on the Planck brane.  We have taken the range $1\leq c\leq
100$, which accomodates for the possibility, \eg, that the third
generation fermions are in the bulk and are localized far from the
Planck brane.
The resulting event rate, in the electron channel only, 
for the nearly degenerate states $Z_{2,3}$ 
is displayed in Fig. \ref{drelly} for the LHC with an integrated 
luminosity of 3 ab$^{-1}$.  This high value of integrated luminosity
corresponds to that proposed for the LHC upgrades \cite{superlhc}.
The apparently isolated single resonance is, of course, a
superposition of the $Z_2$ and $Z_3$ KK states.
The effect of increasing the $Z_n$ width is readily 
visible; the resonance peak becomes flattened if
the total width is too large.  However, it is clear that Drell-Yan 
production provides a clean discovery channel 
for the first two excited states in the 
case $\Gamma_n\lsim 25\Gamma_n^0$.  For present design luminosities,
$\sim 100$ \infb, the event rate is simply scaled by a factor of
30 and the signal remains strong.  The next excitation, $Z_4$, is
very weakly coupled, and we have found that the 
corresponding peak is too small to be observed
above the Drell-Yan SM continuum.  The corresponding event rate 
for the higher mass KK states, $Z_{5,6}$, is also shown in 
Fig. \ref{drelly}, again assuming 3 ab$^{-1}$ of integrated 
luminosity.  Here, we see that the
number of events is small, and even when the $\mu$ channel is also
included these resonances are unlikely to be observed.  Hence
the LHC is likely to only observe a single resonance peak,
corresponding to the superposition of the 
first two $Z$ excitations.  We also expect
that only the first $W$ excitation will be observable at the LHC. 
We note that the visible spectrum of the weak gauge KK states in the WHM 
at the LHC will appear similar to that from a flat extra dimension
with brane terms.  The KK states arising from flat space in the
absence of brane terms will have a larger production rate
\cite{tgrjimbo}, due to
the larger couplings, and will be differentiable from the WHM.

In principle, neutral gauge KK production may be distinguished from
that of more conventional extra gauge bosons arising in, \eg, a GUT model 
\cite{tgrlhclc}.  The presence of the two nearly degenerate KK states
(whether they be the $Z_{2,3}$ in the WHM, or the photon and
$Z$ KK excitations in flat space) results in a unique resonance shape,
which is different from the case of a single new gauge state. 
In the present case, the $Z_2$ and $Z_3$
resonances destructively interfere with the SM background,
yielding the dip in the line-shape in the invariant mass
bins just below the heavy resonances.  This effect is in principle 
measurable at the LHC \cite{gia}, given enough statistics, and is a means for
identifying the production of gauge KK states.  In addition,
the indirect exchange of the $Z_n$ (for $\sqrt s < m_{Z_n}$) 
in fermion pair production
in \epem\ annihilation results in a pattern of deviations in the
cross sections and corresponding asymmetries which allows
for the determination of the fermionic couplings
of additional $Z$ bosons \cite{tgrlc}.  In principle, a TeV class
Linear Collider (LC) could thus be able to resolve the $Z_2$ from
the $Z_3$ and separately measure their couplings.  This claim should be
verified by an independent study.  A multi-TeV LC, such as CLIC,
would be able to run on the resonance peaks, measure the individual
line-shapes, and perform detailed studies of the couplings for
each state.

The KK excitations of the gluon may be produced as resonances in
dijet distributions at the LHC.  The $2\to 2$ parton-level 
subprocesses which contribute to dijet production are $q\bar q\to 
q\bar q\,, q\bar q\to gg\,, qg\to qg\,, gg\to gg\,,$ and $qq\to qq$.
In principle, the gluon KK states can contribute via $s$-channel exchange 
in the $q\bar q$ and $gg$ initiated processes, and via $t$- and $u$-channel
exchange in $qg\to qg$ and $gg\to gg$.  
Here, we are only concerned with the 
search for peaks in the dijet invariant mass distribution, and 
hence neglect the possible gluon KK $t$- and $u$-channel contributions.  
Such contributions would, however, be revealed in dijet angular 
distributions.  We are then left with computing the KK $s$-channel 
exchange diagrams, for which we need to first examine the gluon KK 
couplings to the SM fields.  The expression for the
$q\bar q g_n$ coupling is given in Eq. (60) and its
strength is shown in Fig. \ref{glue} for the first excitation as the brane
kinetic term is allowed to vary.  For the value of the brane terms
present in the WHM, the strength of the square of this coupling is
$0.234\, (g_s^{SM})^2$ for the first gluon excitation and 
$0.143\, (g_s^{SM})^2$
for the second KK mode.  These coupling strengths are a larger
fraction of the usual SM value as compared to the corresponding
couplings of the weak boson KK states due to the large negative value of
$\delta_s$.  Recalling that the zero-mode gluon wavefunctions are 
flat in $z$, it is easy to see that the $g_0g_0g_1$
coupling is forbidden by orthonormality.  Hence the $gg$ initiated
process does not contribute to the resonant production of the KK modes and
the $q\bar q\to q\bar q$ subprocess is the only process we need
to consider here.  We also recall that the gluon KK mass spectrum
tracks that of the $W$ boson KK states with $m_{g_1}=2.53$ TeV
and $m_{g_2}=5.51$ TeV.

\begin{figure}[htbp]
\centerline{
\includegraphics[width=9cm,angle=90]{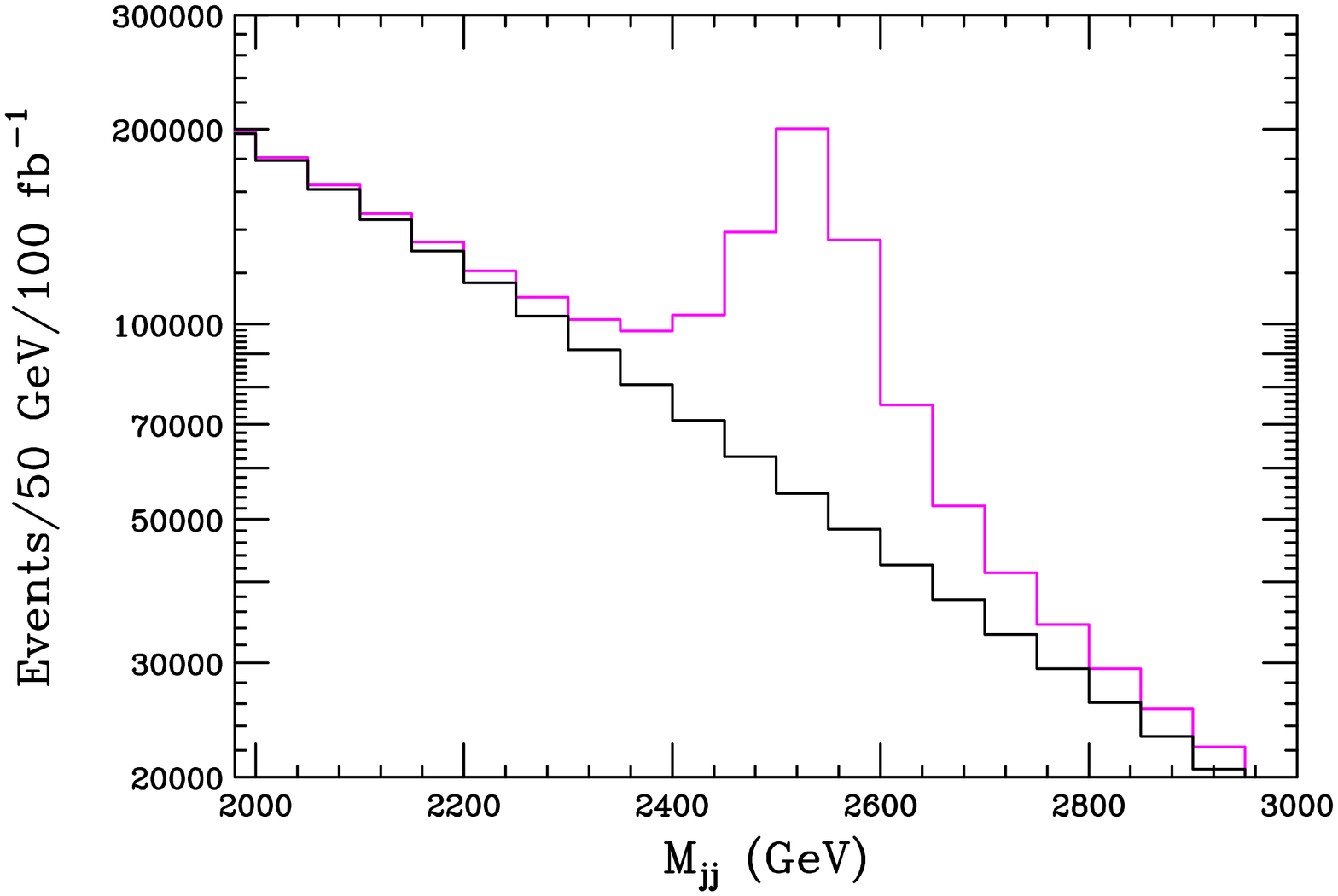}}
\vspace*{0.1cm}
\centerline{
\includegraphics[width=9cm,angle=90]{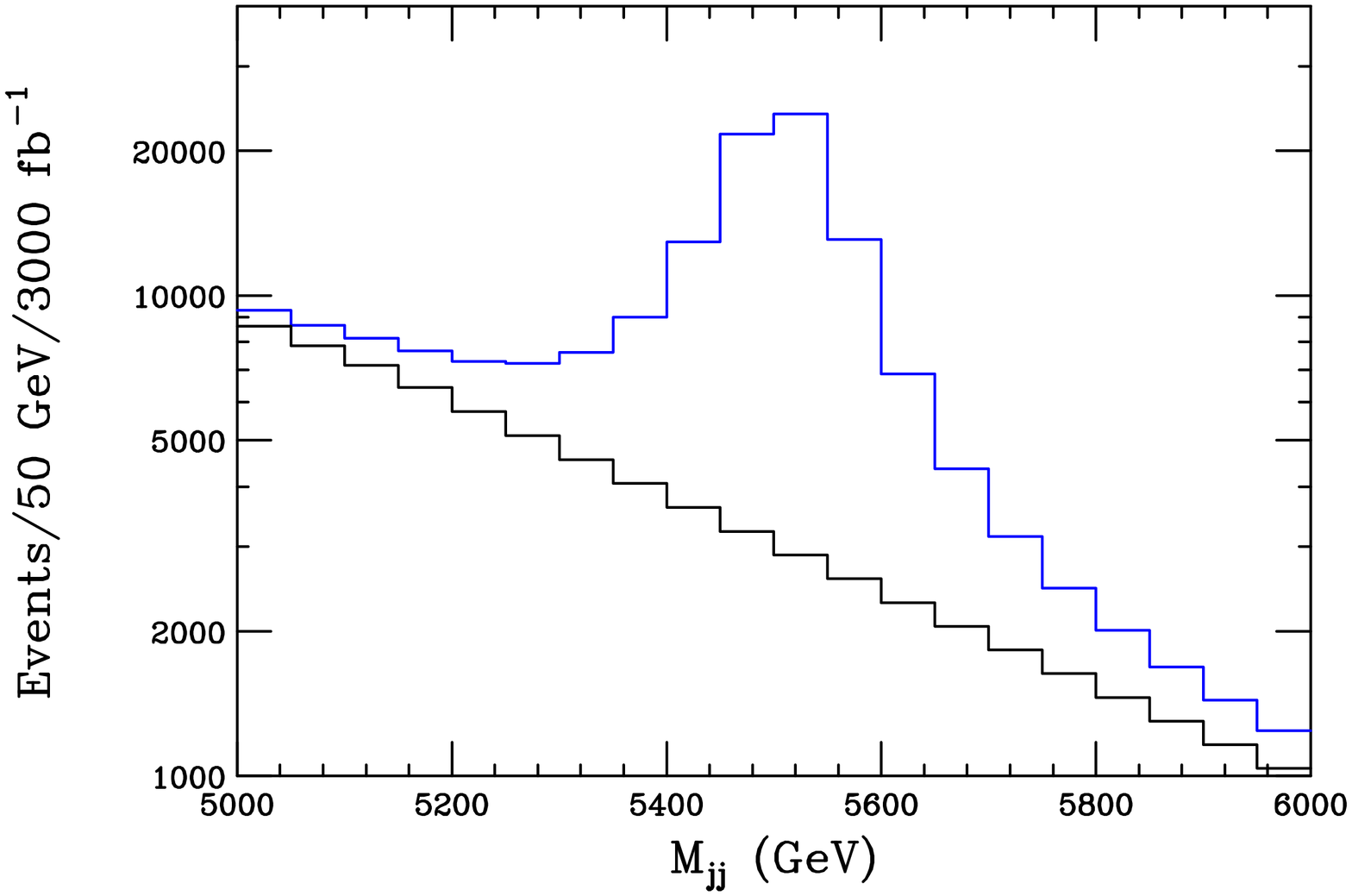}}
\vspace*{0.1cm}
\caption{Production of the first (top panel) and second
(bottom panel) gluon KK excitation in the dijet channel as
a function of the dijet invariant mass. The SM background
is given by the black histogram.}
\label{dijety}
\end{figure}

The resulting event rate for the dijet invariant mass distribution 
is displayed in Fig. \ref{dijety} for the first and second KK 
excitations, taking 100 \infb\ and 3 ab$^{-1}$ of integrated 
luminosity, respectively.  Here,
we have employed the cuts $|\eta|<1$ and $|p_T^{jet_1}|>800 (1500)$ 
GeV for the first (second) excitation.  We see that the event rates
are enormous and that both excitations will be observable at the LHC.
Varying the width, as done above in the case of Drell-Yan production,
will flatten the peak, but should not affect the visibility
of the signal unless the width grows very large.  Observation of
the first dijet resonance, in addition to the peak present in the
Drell-Yan distribution, will signal that the full SM gauge sector
resides in the bulk.  The slightly different value of the mass for
$g_1$, as compared to that for $Z_{2,3}$, with the $g_1$ being
roughly 200 GeV heavier than the $Z$ KK states,
will signal that brane kinetic terms are present in the model.  
Given the large event rate for the production of these KK states, this
mass difference should be measurable.
Observation of the second gluon KK dijet peak will reveal the mass
gap between the states in the KK tower, and will signal the presence 
of a warped, rather than flat, geometry.  Hence, observation of
the KK dijet resonances is critical to the identification of this 
model.

We now turn to the production of the graviton KK states and first
consider the case of resonant graviton production.  This is
well-known to be the main signature of the original RS model
\cite{dhr}.  In principle, resonant graviton production can 
proceed via $q\bar q$ and $gg$ initiated subprocesses.  However,
in the WHM scenario where the fermions
are localized on the Planck brane, the graviton KK tower couples
to fermions with $\mpl^{-1}$ strength or smaller since no warp factor is
generated in the coupling.  Hence the graviton KK tower decouples
from the fermion sector.  Examining the couplings of the graviton 
excitations to the zero-mode vector bosons, we see that in the 
absence of brane terms these are given simply by \cite{dhr3}
\be
g^0_{V_0V_0G_n} = {2\over\Lambda_\pi\pi kr_c}
\left({1-J_0(x_n^G)\over (x_n^G)^2|J_2(x_n^G)|}\right)\,,
\ee
where $x_n^G$ denotes the roots which determine the graviton KK
mass spectrum and, for example, $V_0=g$.
In the presence of brane kinetic terms, both the $V_0$ wavefunction 
and the $V_0V_0G_n$ interaction are modified;
in the case $V_0=g$,
\be 
g_{V_0V_0G_n}={N_{V_0}(\delta_i=0)\over N_{V_0}}\left\{
 g^0_{V_0V_0G_n} + \cdot\cdot\cdot \right\}\,,
\ee
where $\delta_i$ denotes the appropriate brane term.
The omitted terms in the bracket are proportional
to $(x^G_n)^2e^{-2\pi kr_c}$ for $n>0$ and thus are negligible.  Note that
these terms are essential, however, to retain the $\mpl^{-1}$ behavior 
of the zero-mode graviton coupling.  For the case of the graviton
KK tower, the only influence of the brane terms on the $V_0V_0G_n$
coupling arises from modifications of the vector boson wavefunction.

\begin{figure}[htbp]
\centerline{
\includegraphics[width=9cm,angle=90]{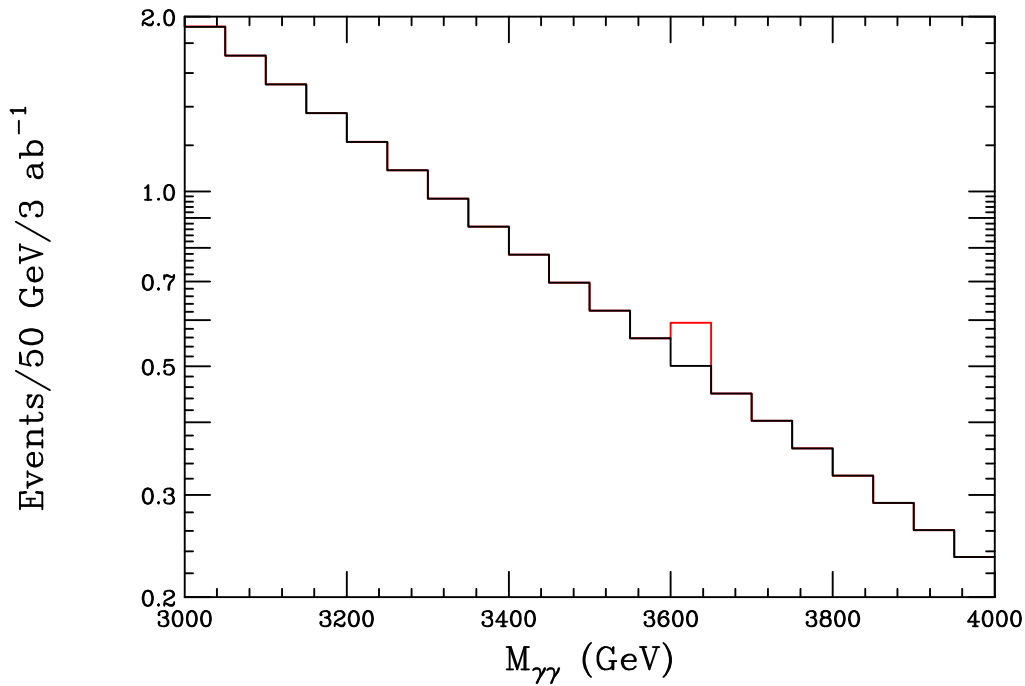}}
\vspace*{0.1cm}
\caption{Production rate for the first graviton excitation at
the LHC via the process $gg\to G_1\to\gamma\gamma$ as a
function of the diphoton invariant mass.  The SM diphoton
background is also shown.  The two histograms are indistinguishable
except for the small blip at $M_{\gamma\gamma}=m_{G_1}$ }
\label{diphot}
\end{figure}

Resonant graviton KK production thus proceeds through 
$gg\to G_n\to\gamma\gamma\,, gg\,, ZZ\,, WW$.  Since the $G_n$ coupling
is significantly weaker than that for the gluon excitations, we
expect that the $gg$ channel and $ZZ,WW$ decay to 
hadronic final states will be
overwhelmed by the SM background.  Likewise, we expect 
the rate for the leptonic final states to be small due to the
low $ZZ,WW$ leptonic branching fractions.  Thus, we only consider 
the $\gamma\gamma$ final state.  The SM diphoton background 
arises from $q\bar q\to\gamma\gamma$ and $gg\to\gamma\gamma$,
where the latter process proceeds through a box diagram.  We
include both of these SM processes in our background calculation.
The event rate at the LHC, with 3 ab$^{-1}$ of integrated
luminosity, is displayed in Fig. \ref{diphot} 
for the first graviton excitation and
the SM background as a function of the diphoton invariant mass.
In our numerical calculations, we assume $k/\mpl=0.1$.
We see that the $G_1$ production has a very small event rate and 
is indistinguishable from
the background.  Hence, the WHM differs from the usual RS
scenario in that graviton resonances will not be observed.

For completeness, we also considered the associated production 
of KK gravitons via $gg\to G_n+g$.   Appropriately modifying
the expressions in Ref. \cite{grw} for the WHM we computed the
event rate at the LHC for $G_1$ production 
as a function of jet energy using an integrated luminosity of
3 ab$^{-1}$.  We found that for typical jet energies of $E_j=200$ GeV the
cross section was of order 0.016 ab, and
hence is also too small to be observed, even with the
proposed LHC luminosity upgrades.  We thus conclude that in
this model, the graviton KK tower can not be observed at high 
energy colliders.

Lastly, we note that there exists a radion scalar in this model
and that it may have distinctive collider signatures.

\section{Conclusions}

Various phenomenological aspects of a Higgsless 5-d
model \cite{CsakiII}, based on the RS hierarchy proposal \cite{RS},
were studied in this paper.  We considered independent left
and right bulk gauge
couplings and included the effects of UV brane localized kinetic
terms for the gauge fields \cite{nomura}.  These terms were assumed
to be radiatively generated, which is a generic expectation in
orbifold  models \cite{georgi}.  Our analysis was not limited to
leading order bulk-curvature effects unlike in
Refs. \cite{CsakiII,nomura}, and also allowed for a more general
set of parameters than that discussed in Ref. \cite{Barbi}.

We computed the mass spectrum and the relevant couplings of the
$W^\pm$ and $\gamma/Z$ KK towers, and studied experimental
constraints on the model parameters.  Our main conclusion is that in
the region of parameter space allowed by precision EW data, this
model is not perturbatively 
unitary at tree level above $\sqrt{s} \approx 2$ TeV,
which is below the scale of the new KK states. 
Futhermore, we find that tree-level unitarity is violated over the entire 
parameter space, even in those
regions where comparisons with the precision measurements are
anticipated to be quite poor.  Thus, to make
reliable calculations based on the WHM, one must extend this model
in order to unitarize the amplitudes. Setting the issue of perturbative
unitarity aside, it was also observed that quantum contributions to
the $S$, $T$, $U$ oblique parameters \cite{PT,obstu} are expected to be
small. However, in the absence of the Higgs, regularization of the
relevant loop diagrams may require non-renormalizable TeV brane
counter terms whose coefficients are unknown.  This imposes a
degree of uncertainty on loop corrections.  Further work regarding
loop corrections is needed before more precise statements could be
made in this regard.

Finally, we considered the collider signatures of the model,
assuming that unitarity could somehow be restored without
significantly modifying our numerical results. These signatures
depend on the 5-d configuration of bulk fermions. We assumed a
simple setup, where all fermions, except perhaps for the third
generation, are localized near the Planck brane.  The effect of
different localizations of quarks was then taken into account by
varying the widths of the KK resonances. Generically, we found
that the low-lying gauge boson KK modes, including the gluons,
would be observable, whereas the most distinct RS signature, the
spin-2 graviton KK resonances, would most likely evade detection
at the LHC.

The AdS/CFT correspondence \cite{Maldacena} provides a 4-d
interpretation of this model in terms of strong dynamics. Thus,
the tools and insights of both five and four dimensional model building can
be employed in making this scenario more realistic such that it
agrees with the SM at low energies.  This setup provides an
entirely higher dimensional explanation of the observed weak
interaction mass scales, directly linking them to the IR scale in
the RS model.  Thus, it is worth the effort to find solutions for
the problems that plague the present form of the WHM.

\noindent{\bf Acknowledgements}  
The work of H.D. was supported by the Department of Energy under
grant DE-FG02-90ER40542.
We would like to thank Nima Arkani-Hamed, 
Tim Barklow, Graham Kribs, Hitoshi Murayama, and Michael Peskin for
discussions related to this work.

\end{document}